\documentclass[
  journal=largetwo,
  manuscript=article-type,
  year=2020,
  volume=37,
]{cup-journal}

\usepackage{amsmath}
\usepackage[nopatch]{microtype}
\usepackage{booktabs}
\usepackage{orcidlink}
\usepackage{amsfonts}
\usepackage{amssymb}
\usepackage{esint}
\usepackage{bm}
\usepackage{multirow}
\usepackage{makecell}

\DeclareUnicodeCharacter{0301}{\'{}}

\title{Forecast Measurement of the 21\,cm Global Spectrum from Lunar Orbit with the Vari-Zeroth-Order Polynomial (VZOP) Method}

\author{Tianyang Liu\,\orcidlink{0009-0001-0228-9130}}
\affiliation{Shanghai Astronomical Observatory, Chinese Academy of Sciences, Shanghai 200030, China}
\alsoaffiliation{School of Astronomy and Space Science, University of Chinese Academy of Sciences, Beijing 100049, China}

\author{Jiajun Zhang\,\orcidlink{0000-0002-4117-343X}}
\affiliation{Shanghai Astronomical Observatory, Chinese Academy of Sciences, Shanghai 200030, China}
\email[Jiajun Zhang \& Junhua Gu \& Quan Guo]{jjzhang@shao.ac.cn, jhgu@nao.cas.cn, guoquan@shao.ac.cn}

\author{Yuan Shi\,\orcidlink{0000-0001-8233-3703}}
\affiliation{Department of Astronomy, School of Physics and Astronomy, Shanghai Jiao Tong University, Shanghai 200240, China}
\alsoaffiliation{Key Laboratory for Particle Astrophysics and Cosmology (MOE) / Shanghai Key Laboratory for Particle Physics and Cosmology, China}

\author{Junhua Gu\,\orcidlink{0000-0001-9765-6521}}
\affiliation{National Astronomical Observatories, Chinese Academy of Sciences, Beijing 100101, China}

\author{Quan Guo\,\orcidlink{0000-0003-2858-5090}}
\affiliation{Shanghai Astronomical Observatory, Chinese Academy of Sciences, Shanghai 200030, China}

\author{Yidong Xu\,\orcidlink{0000-0003-3224-4125}}
\affiliation{National Astronomical Observatories, Chinese Academy of Sciences, Beijing 100101, China}
\alsoaffiliation{Key Laboratory of Radio Astronomy and Technology, Chinese Academy of Sciences, A20 Datun Road, Chaoyang District, Beijing, 100101, P. R. China}

\author{Furen Deng\,\orcidlink{0000-0001-8075-0909}}
\affiliation{School of Astronomy and Space Science, University of Chinese Academy of Sciences, Beijing 100049, China}
\alsoaffiliation{National Astronomical Observatories, Chinese Academy of Sciences, Beijing 100101, China}

\author{Fengquan Wu\,\orcidlink{0000-0002-6174-8640}}
\affiliation{National Astronomical Observatories, Chinese Academy of Sciences, Beijing 100101, China}

\author{Yanping Cong\,\orcidlink{0000-0002-4456-6458}}
\affiliation{Shanghai Astronomical Observatory, Chinese Academy of Sciences, Shanghai 200030, China}

\author{Xuelei Chen\,\orcidlink{0000-0001-6475-8863}}
\affiliation{School of Astronomy and Space Science, University of Chinese Academy of Sciences, Beijing 100049, China}
\alsoaffiliation{National Astronomical Observatories, Chinese Academy of Sciences, Beijing 100101, China}
\alsoaffiliation{Department of Physics, College of Sciences, Northeastern University, Shenyang 110819, China}
\alsoaffiliation{Center for High Energy Physics, Peking University, Beijing 100871, China}

 %% First letter not capped

\begin{document}

\begin{abstract}
The cosmic 21\,cm signal serves as a crucial probe for studying the evolutionary history of the Universe. However, detecting the 21\,cm signal poses significant challenges due to its extremely faint nature. To mitigate the interference from the Earth's radio frequency interference (RFI), the ground and the ionospheric effects, the Discovering the Sky at the Longest Wavelength (DSL) project will deploy a constellation of satellites in Lunar orbit, with its high-frequency daughter satellite tasked with detecting the global 21\,cm signal from cosmic dawn and reionization era (CD/EoR). We intend to employ the Vari-Zeroth-Order Polynomial (VZOP) for foreground fitting and subtracting. We have studied the effect of thermal noise, thermal radiation from the Moon, the Lunar reflection, anisotropic frequency-dependent beam, inaccurate antenna beam pattern, and RFI contamination.  We discovered that the RFI contamination can significantly affect the fitting process and thus prevent us from detecting the signal. Therefore, experimenting on the far side of the moon is crucial. We also discovered that using VZOP together with DSL, after 1080 orbits around the Moon, which takes about 103 days, we can successfully detect the CD/EoR 21\,cm signal.
\end{abstract}

\section{Introduction}\label{sec:intro}
The Cosmic Dawn (CD) typically refers to the epoch when the first luminous objects began to form. After a prolonged Dark Age, the initial generation of luminous objects gradually emerged through gravitational collapse, re-illuminating and warming the Universe. Subsequently, the ultraviolet and X-ray emissions from these first stars heated and ionized the surrounding neutral hydrogen, marking the era known as the Epoch of Reionization (EoR), as the Universe was initially ionized after the Big Bang. Research on the CD and EoR is actively ongoing, with the most promising probe currently believed to be the HI 21\,cm signal. The 21\,cm signal originates from the hyperfine splitting of the 1S ground state of hydrogen, with a wavelength of 21\,cm in the rest frame (1.42\,GHz). During the CD and EoR, the expansion of the Universe and the heating by the first generation of celestial objects caused a divergence between the gas and the CMB temperature over time \citep[][]{pritchard201221}. The interactions between HI and matter or radiation resulted in distinct absorption or emission features at different times. Due to the expansion of the Universe, features from various periods have been redshifted to various frequencies, transforming the original line spectrum into the current continuous spectrum. The cosmic 21\,cm signal from the CD and EoR is generally believed to originate from the period corresponding to redshifts in the range of 6-27, placing it within the meter-wave band, with a frequency range of approximately 50-200\,MHz. The spectrum of the cosmic 21\,cm signal reflects the interactions between matter and radiation during the CD and EoR. This provides us with additional information about the formation and evolution of the Universe, serving as a key piece of the puzzle in reconstructing the history of cosmic evolution.

However, detecting the cosmic 21\,cm signal faces significant challenges, primarily because it is exceedingly faint, expecting less than 200\,mK \citep[][]{cohen2018charting, reis2021subtlety, xu2021maximum}. The foreground temperature generated by synchrotron radiation in the Galaxy, in contrast, surpasses the signal by approximately 4-5 orders of magnitude. Beyond this, various other contaminations such as extragalactic sources, deviations caused by ionospheric absorption and refraction, radio frequency interference (RFI), ground reflections, and noise from the instruments themselves increase the difficulties. The noise temperature caused by these noise sources is orders of magnitude greater than the 21\,cm signal.

While detecting the 21\,cm signal remains challenging, numerous ground-based experiments are currently underway. These experiments generally follow four strategies: (1) conducting global signal observations using a single antenna, (2) employing antenna arrays for power spectrum measurements to study statistical properties of the 21\,cm signal, (3) directly performing imaging observations - a key scientific objective for the upcoming SKA \citep[Square Kilometre Array;][]{mellema2013reionization, koopmans2015cosmic, bolli2020test}, and (4) observing 21\,cm forest, which uses the distant quasars as background sources instead of the CMB \citep[][]{carilli2002hi, xu200921, xu2011earliest, shao202321}. Among these, global signal measurements are widely implemented due to the simplicity of single-antenna design, cost-effectiveness, and portability. Some well-known global experiments include EDGES \citep[Experiment to Detect the Global EOR Signature;][]{rogers2012absolute, bowman2018absorption, murray2022bayesian, rogers2022analytic}, SARAS3 \citep[Shaped Antenna measurement of the background RAdio Spectrum 3;][]{girish2020saras, subrahmanyan2021saras}, REACH \citep[Radio Experiment for the Analysis of Cosmic Hydrogen;][]{de2019reach, de2022reach, cumner2022radio, kirkham2024bayesian}, LEDA \citep[Large-Aperture Experiment to detect the Dark Ages;][]{greenhili2012broadband, garsden202121, spinelli2022antenna}, BIGHORNS \citep[Broadband Instrument for Global HydrOgen ReioNisation Signal;][]{sokolowski2015bighorns, sokolowski2015impact}, SCI-HI \citep[Sonda Cosmológica de las Islas para la Detección de Hidrógeno Neutro;][]{voytek2014probing}, and PRI$^Z$M \citep[Probing Radio Intensity at high-Z from Marion;][]{philip2019probing}. Among these experiments, only EDGES claims to have successfully detected the 21\,cm signal, which manifests as a flattened absorption profile with a depth of more than 500\,mK \citep[][]{bowman2018absorption}. However, due to the signal's depth being several times greater than theoretically expected, it has not been widely accepted \citep[][]{cohen2018charting}. SARAS has investigated this profile and suggests that it is a spectral distortion caused by noise in the low-frequency instrument of EDGES \citep[][]{singh2022detection}.

Detecting the EoR signal from the Earth's surface presents a significant challenge. Due to the dielectric properties of the soil, standing waves occur between the ground and the antenna, leading to sinusoidal systematic errors in antenna temperature \citep[][]{bevins2021maxsmooth, bevins2022comprehensive}. The Earth's ionosphere strongly refracts, reflects, and absorbs low-frequency electromagnetic waves, the deviation which induces in antenna temperature is 2$\sim$3 orders of magnitude larger than the global 21\,cm signal \citep[][]{vedantham2014chromatic, shen2021quantifying, wang2024tackling}. Additionally, there is a substantial amount of natural and artificial radio emissions within this frequency range on Earth. Due to Sporadic E propagation or Tropospheric propagation, the observation of the 21\,cm signal is further disrupted \citep[][]{sokolowski2015impact}. The noise generated by any of the aforementioned effects may be much larger than the amplitude of the 21\,cm signal.

To mitigate various interferences, since the onset of the space age, there has been a pursuit to locate a place beyond Earth's confines that is free from radio interference for astronomical observations. The far side of the Moon has emerged as an ideal candidate for this purpose. The Moon has no atmosphere, and therefore, no ionosphere. Moreover, the Lunar far side effectively shields against Earth's electromagnetic radiation and, during its nighttime, concurrently screens out the Solar low-frequency radiation. Currently, there are several planned space and Lunar-based experiments, aimed at effectively avoiding interference from Earth's RFI and the ionosphere \citep[][]{silk2021astronomy, koopmans2021peering}, such as DSL \citep[Discovering the Sky at the Longest Wavelength, or \emph{Hongmeng} in Chinese;][]{boonstra2016discovering, chen2019discovering, chen2021discovering, chen2023discovering}, DAPPER \citep[Dark Ages Polarimeter PathfindER;][]{tauscher2018new, burns2021transformative} and its precursor DARE \citep[Dark Ages Radio Explorer;][]{burns2017space}, and FARSIDE \citep[Farside Array for Radio Science Investigations of the Dark ages and Exoplanets;][]{burns2019nasa, burns2021lunar, burns2021transformative}.

The Lunar far side offers a tranquil radio environment, yet contaminants from cosmic sources remain a pressing challenge, with the Galactic foreground being the most contaminant. Within our target frequency range, the Galactic foreground exhibits a brightness temperature as high as approximately $\sim10^3$\,K, whereas the deepest point of the cosmic 21\,cm signal absorption trough is only about $\sim10^{-1}$\,K. Before this, the systematic errors of the antenna itself must be calibrated \citep[][]{sun2024calibration}. Common foreground subtraction methods typically involve fitting a polynomial to model the foreground \citep[][]{mozdzen2016limits, bernardi2016bayesian, gu2020direct, shi2022lunar, singh2022detection, tripathi2024extracting}. This approach was employed by EDGES in the discovery of the 21\,cm signal \citep[][]{bowman2018absorption}. The approach stems from the fact that within the frequency range where the sharp 21\,cm signal resides during the EoR, the Galactic synchrotron radiation dominates and manifests as a smooth power-law spectrum \citep[][]{shaver1999can}. Unfortunately, due to antenna chromaticity, it couples the spatial structure of the Galactic foreground in the frequency domain. Consequently, the antenna temperature obtained is not a perfect power-law spectrum, leading to poor performance of polynomial fitting. Researchers attempted to correct the distortion of the total power spectrum using beam correction factors \citep[][]{sims2020testing, shen2021quantifying, anstey2022informing, de2022reach, spinelli2022antenna}; however, the effectiveness of this approach was limited. In theory, if we have knowledge of the antenna beam pattern, we can invert the additional spectral structure introduced by antenna chromaticity. Based on this idea, we have proposed an improved polynomial fitting algorithm called VZOP \citep[Vari-Zeroth-Order Polynomial,][]{liu2024detecting}.

In \citet{liu2024detecting}, we confirmed the capability of VZOP to recover the 21\,cm signal accurately from foregrounds. Therefore, this paper aims to apply VZOP in the upcoming DSL project. The rest of this paper is structured as follows: Section~\ref{sec:DSL} provides a brief overview of the DSL project and Section~\ref{sec:VZOP_algortithm} introduces the VZOP algorithm with appropriate modifications. Section~\ref{sec:Simulation} describes the foreground model, cosmological signal model, antenna, and thermal noise model used in this paper, along with the simulation process. We present the performance of VZOP under ideal conditions in Section~\ref{sec:Results}. In Section~\ref{sec:NIS}, we present the results under non-ideal conditions, examining the impact of various errors on observations. Finally, we discuss and conclude in Section~\ref{sec:DAC}.

\section{DSL Project}\label{sec:DSL}
The DSL project is envisaged as a formation of 1 main satellite and 9 daughter satellites orbiting the Moon, forming a space-distributed interferometric array with baselines ranging from 100\,m to 100\,km. Among these, one high-frequency daughter satellite will operate within the 30-120\,MHz range, primarily conducting global cosmic 21\,cm signal measurements. Additionally, the remaining 8 daughter satellites will primarily operate in the ultra-long wavelength band of 0.1-30\,MHz, performing interferometric measurements while also being capable of conducting spectral measurements. The high-frequency daughter satellite employs a sleeve loaded spherical cone antenna (ice cream antenna hereafter), whereas the low-frequency daughter satellites employ dipole antennas. To maintain a stable flying formation, each satellite is oriented such that one side continuously faces the centre of the Moon. The focus of this study is on the cosmic 21\,cm signal's absorption trough in the frequency range of 50-120\,MHz, corresponding to a redshift of approximately $z=11-27$. Consequently, our discussion primarily revolves around the high-frequency daughter satellite, which does not engage in interferometric measurements with other daughter satellites.

\section{VZOP Algorithm}\label{sec:VZOP_algortithm}
In previous work, we proposed the VZOP algorithm, which incorporates antenna beam pattern information into a polynomial fitting model \citep[][]{liu2024detecting}. This approach ultimately fits the antenna temperature spectrum to a pseudo-polynomial model with a zeroth-order term that varies with frequency. We included the antenna beam information in the zero-order term, incorporating the knowledge of the antenna beam pattern into the model to invert the additional spectral structure caused by antenna chromaticity.

We have introduced a 24-hour averaged beam model into VZOP. For the symmetrical ice cream antenna that may be placed on the high-frequency daughter satellite, VZOP can be naturally applied. However, VZOP is, in fact, a general method that can also be employed with non-symmetrical antennas. In that case, the satellite must have a rotation in theory. The high-frequency daughter satellite does not participate in interferometer measurements with other satellites, which allows for the independent design of its flight attitude. We can set the high-frequency daughter satellite (and hence its antenna) to undergo self-rotation about an axis perpendicular to the Lunar surface, with the shortest possible rotation period.

Then, in this paper, we need to make some minor modifications to VZOP to facilitate its application to non-axisymmetric antennas. First, we establish a coordinate system that moves along with the satellite but does not rotate with it. In this coordinate system, the zenith direction always points in the opposite direction to the Lunar centre, where $\theta$ represents the polar angle, and $\phi$ is the azimuthal angle, with $0^\circ$ indicating the direction of the satellite's motion. We denote $p$ as the satellite's self-rotation period, and the antenna temperature $\bar{T}_A(\nu)$ (In our simulation, we consider the integrated sky temperature $T_\mathrm{sky}$, including noise, as the antenna temperature) obtained after integrating over $n_0$ periods is given by:
\begin{equation}
	\bar{T}_A(\nu)=\dfrac{1}{n_0p}\sum_{n=1}^{n_0}\int_{t_n-p/2}^{t_n+p/2}{\left[\oiint{B(\nu,\mathbf{n},t)\mathcal{S}(\mathbf{n})T_b(\nu,\mathbf{n},t)}\mathrm{d}\Omega\right]}\mathrm{d}t,
	\label{eq:antenna_temperature_origin}
\end{equation}
where $T_b(\nu,\mathbf{n},t)$ represents the real sky temperature at frequency $\nu$, direction $\mathbf{n}$, and time $t$. $\mathcal{S}(\mathbf{n})$ is the shade function that represents the shadowing effect caused by the Moon, taking the value of 0 where the Moon blocks and 1 elsewhere. And $B(\nu,\mathbf{n},t)$ corresponds to the antenna gain and $\Omega$ is the solid angle. Here, $B(\nu,\mathbf{n},t)$ is the integral normalized antenna power pattern, but due to the Lunar obstruction covering a portion of the antenna's field of view, the effective beam $B(\nu,\mathbf{n}, t)\mathcal{S}(\mathbf{n})$ integrates over the entire sky to a value less than unity. Actually, the Moon emits radiation, including intrinsic emission and reflection. This aspect is discussed later in Section~\ref{subsec:simulation_process}, and most of the simulation results in this study account for Lunar radiation. However, it is unnecessary to consider this here, as the purpose of this section is to revisit and refine the derivation of VZOP. When using VZOP as a fitting model, we consistently assume that the Moon emits no radiation, i.e., the portion of $\mathcal{S}(\mathbf{n})$ obscured by the Moon is always set to zero. The impact of Lunar radiation is treated as an additional noise source in the antenna temperature spectrum.

Since the reference frame continuously moves along with the satellite, the zenith direction in this coordinate system undergoes variations as the satellite orbits the Moon. Consequently, the sky temperature map in this frame also experiences changes. If the antenna's self-rotation period is short, these variations are relatively minor. Therefore, we can express equation~(\ref{eq:antenna_temperature_origin}) in the following form:
\begin{align}
	\bar{T}_A(\nu)\!
	&\approx
	\dfrac{\!\sum_{n=1}^{n_0}\!\iiint{\!\!\!B(\nu,\theta,\phi,t_n)\mathcal{S}(\theta)T_b(\nu,\theta,\phi-\phi',t_n)\cos{\theta}\mathrm{d}\theta\mathrm{d}\phi\mathrm{d}\phi'}}{2{\pi}n_0} \nonumber\\
	&=\dfrac{\sum_{n=1}^{n_0}\int{\mathcal{S}(\theta)\bar{T}_b(\nu,\theta,t_n)\bar{B}(\nu,\theta,t_n)\cos{\theta}\mathrm{d}\theta}}{n_0},
\end{align}
where $\bar{B}(\nu,\theta,t_n)$ is defined as the beam model averaged along the azimuthal angle at time $t_n$, $\bar{T}_b(\nu,\theta,t_n)$ is defined as the sky temperature model averaged along the azimuthal angle at time $t_n$, and shade function $\mathcal{S}(\theta)$ is independent of $\phi$. Because the coordinate system is in motion, $\bar{T}_b(\nu,\theta,t_n)$ is dependent on $t_n$, while $\bar{B}(\nu,\theta,t_n)$ is independent of $t_n$. By interchanging the order of summation and integration, we obtain the following expression:
\begin{align}
    \bar{T}_A(\nu)
	&=\int{\mathcal{S}(\theta)\sum_{n=1}^{n_0}\dfrac{\bar{T}_b(\nu,\theta,t_n)}{n_0}\bar{B}(\nu,\theta)\cos{\theta}\mathrm{d}\theta} \nonumber\\
	&=\int{\mathcal{S}(\theta)\bar{T}_b(\nu,\theta)\bar{B}(\nu,\theta)\cos{\theta}\mathrm{d}\theta},
\end{align}
where $\bar{T}_b(\nu,\theta)$ is defined as the time-averaged mean sky temperature model, and $\bar{B}(\nu,\theta)$ is the mean beam model. We discretize both frequency and polar angle:
\begin{equation}
    \bar{T}_A(\nu_i)\approx\sum_j\mathcal{S}(\theta_j)\bar{T}_b(\nu_i,\theta_j)\bar{B}(\nu_i,\theta_j)\cos{\theta_j}.
\end{equation}

So far, we have redefined the mean beam model and the mean sky temperature model. The remaining derivations can be found in Equations (8) to (12) as presented in \citet{liu2024detecting}. In this paper, we evenly divide the elevation angle range $[0, \Theta_m]$ into $N_\mathrm{bin}$ elevation bins, where $\Theta_m$ represents a maximum zenith angle, beyond which the sky is obscured by the Moon. In this study, we employ a 5th-order VZOP to fit the simulated data, unless otherwise stated, with $N_\mathrm{bin}$ set to $10$.

To distinguish them from the simulated data, we use a hat symbol to denote values obtained through sampling with VZOP. Like the Equation (13) in \citet{liu2024detecting}, we employ the same truncated likelihood as the posterior:
\begin{multline}
	\log p(\mathbf{a},\hat{\mathbf{T}}_\mathrm{gal},\mathbf{p}_\mathrm{eor}\vert\mathbf{T}_A)\propto -\dfrac{1}{2}\cdot \\
	\left[\left(\mathbf{T}_A\!-\!\mathbf{S}(\mathbf{a})\mathbf{B}\hat{\mathbf{T}}_\mathrm{gal}\!-\!\hat{\mathbf{T}}_\mathrm{eor}\right)^T\!\!\bm{\Sigma}^{-1}\!\!\left(\mathbf{T}_A\!-\!\mathbf{S}(\mathbf{a})\mathbf{B}\hat{\mathbf{T}}_\mathrm{gal}\!-\!\hat{\mathbf{T}}_\mathrm{eor}\right)\right].
	\label{eq:likelihood}
\end{multline}
The bold characters represent matrices, for example, $\mathbf{B}$ is the mean beam matrix, where the number of rows equals the number of frequency channels, and the number of columns equals the number of elevation bins. In equation~(\ref{eq:likelihood}), the shade function $\mathcal{S}$ has been omitted, and $\hat{\mathbf{T}}_\mathrm{gal}$ represents the foreground temperature in the region not shielded by the Moon. These matrices are specifically defined as follows:
\begin{equation*}
	\begin{cases}
		B_{i,j}\equiv\bar{B}(\nu_i,\theta_j)\cos{\theta_j}, \\
		\mathbf{S}(\mathbf{a})=\mathrm{diag}[S(\nu_1;\nu_r,\mathbf{a}),S(\nu_2;\nu_r,\mathbf{a})\ldots], \\
		\mathbf{T}_A=[T_A(\nu_1), T_A(\nu_2)\ldots]^T, \\
		\mathbf{T}_\mathrm{gal}=[T_\mathrm{gal}(\nu_r,\theta_1), T_\mathrm{gal}(\nu_r,\theta_2)\ldots]^T, \\
		\mathbf{T}_\mathrm{eor}=[T_\mathrm{eor}(\nu_1), T_\mathrm{eor}(\nu_2)\ldots]^T. \\
	\end{cases}
\end{equation*}
where $\nu_r$ is the central frequency, and $S(\nu_i;\nu_r,\mathbf{a})$ contains the non-zeroth order terms in the polynomial:
\begin{equation}
	S(\nu_i;\nu_r,\mathbf{a})\equiv \exp\left[\sum_{n=1}^Na_n\log^n(\dfrac{\nu_i}{\nu_r})\right].
	\label{eq:polynomial}
\end{equation}
VZOP incorporates antenna beam-related information into the model to fit the foreground, thereby enabling the removal of additional spectral structures caused by the chromaticity of the beam.

\section{Simulation}\label{sec:Simulation}
To prepare for forthcoming observations, this section employs the existing sky brightness temperature model, 21\,cm signal model, antenna models, and noise model for simulated observations in Lunar orbit. During the simulation, we also consider scenarios involving Lunar radiation. While Lunar radiation has received minimal discussion previously due to its relatively low intensity, its brightness temperature compared to the 21\,cm signal is non-negligible.

\subsection{Foreground model}
We use the ULSA (Ultra-Long wavelength Sky model with Absorption) model as our foreground model \citep[][]{cong2021ultralong}. The ULSA model is a Python code package developed to generate sky maps at very low frequencies, specifically in the ultra-long wavelength or ultra-low frequency band. This model takes into account the free-free absorption effect caused by free electrons in both discrete HII clumps and the Warm Ionized Medium (WIM). This is its greatest advantage over other foreground models, such as GSM \citep[Galactic Global Sky Model;][]{de2008model}, as it is closer to reality by considering free-free absorption. The ULSA model is designed to provide a reasonable estimate of the sky brightness at ultra-low frequencies, which is crucial for designing low-frequency experiments and interpreting observations in this unexplored part of the electromagnetic spectrum. The model can be used for end-to-end simulations in experiment design and real observations with limited antenna elements. Overall, the ULSA model aims to provide a comprehensive and accurate full-sky model for ultra-low frequency observations, taking into account the significant effects of free-free absorption and providing valuable information for the study of the Galactic and extragalactic radio background.

\begin{figure}
    \centering
	\includegraphics[width=1.\columnwidth]{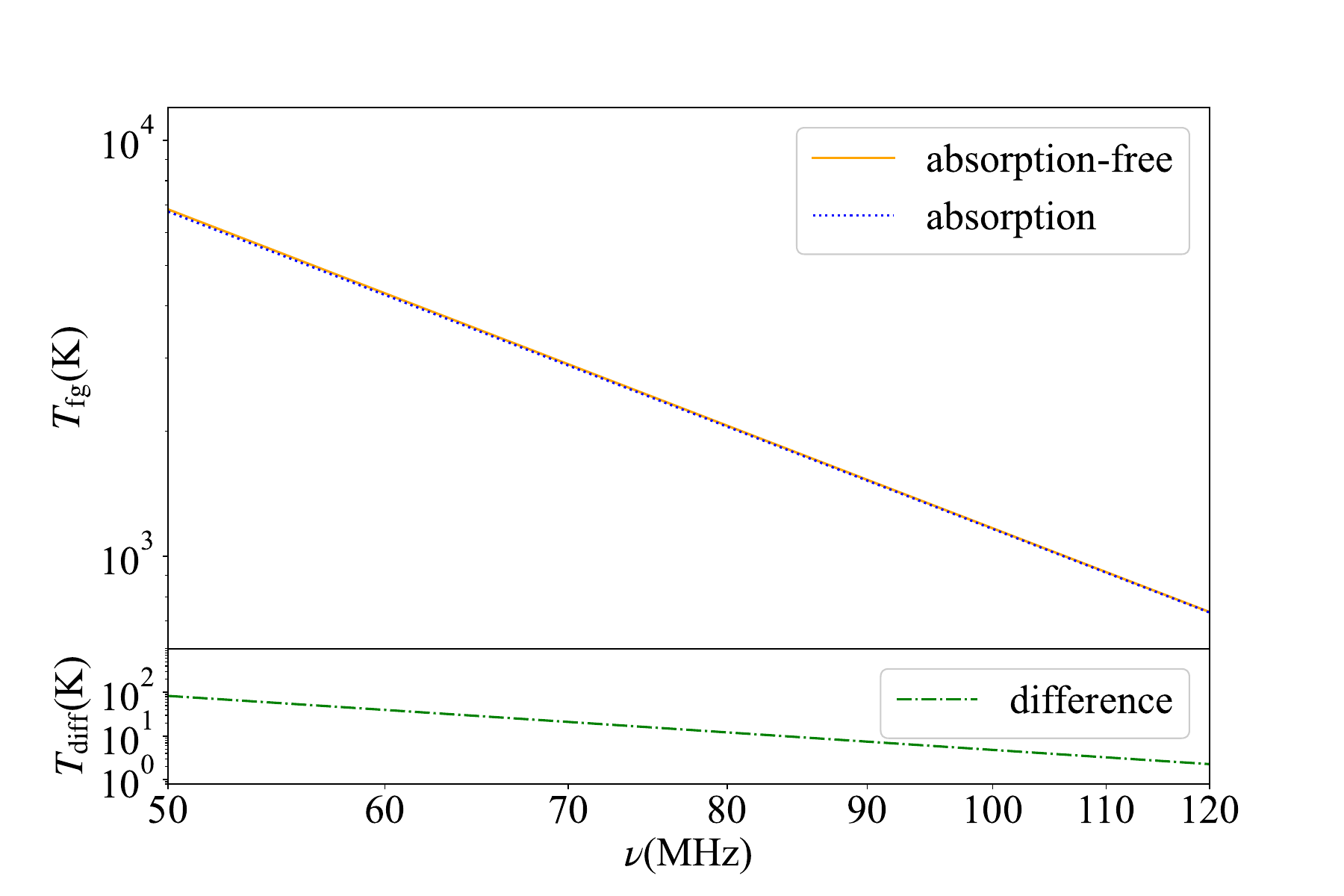}
	\caption{The top panel shows the foreground global spectrum based on the ULSA model using the direction-dependent spectral index. The solid orange line represents the case without considering absorption, while the dotted blue line represents the case accounting for absorption. The bottom panel shows the difference between the two cases.}
	\label{fig:foreground_comparison}
\end{figure}

Fig.~\ref{fig:foreground_comparison} illustrates the integrated global spectrum based on the ULSA model using the direction-dependent spectral index. The upper panel presents the spectra with and without absorption effects, displaying clear power-law characteristics. The lower panel shows the difference between the two, indicating a decrease in antenna temperature by of order 10\,K due to absorption effects within the 50-120\,MHz range, also following a power-law spectrum. It can be speculated that absorption effects will not significantly affect foreground fitting at this frequency range.

\subsection{Cosmic 21\texorpdfstring{$\,$}{}cm signal model}
\label{subsec:signal_model}
\begin{figure}
	\includegraphics[width=\columnwidth]{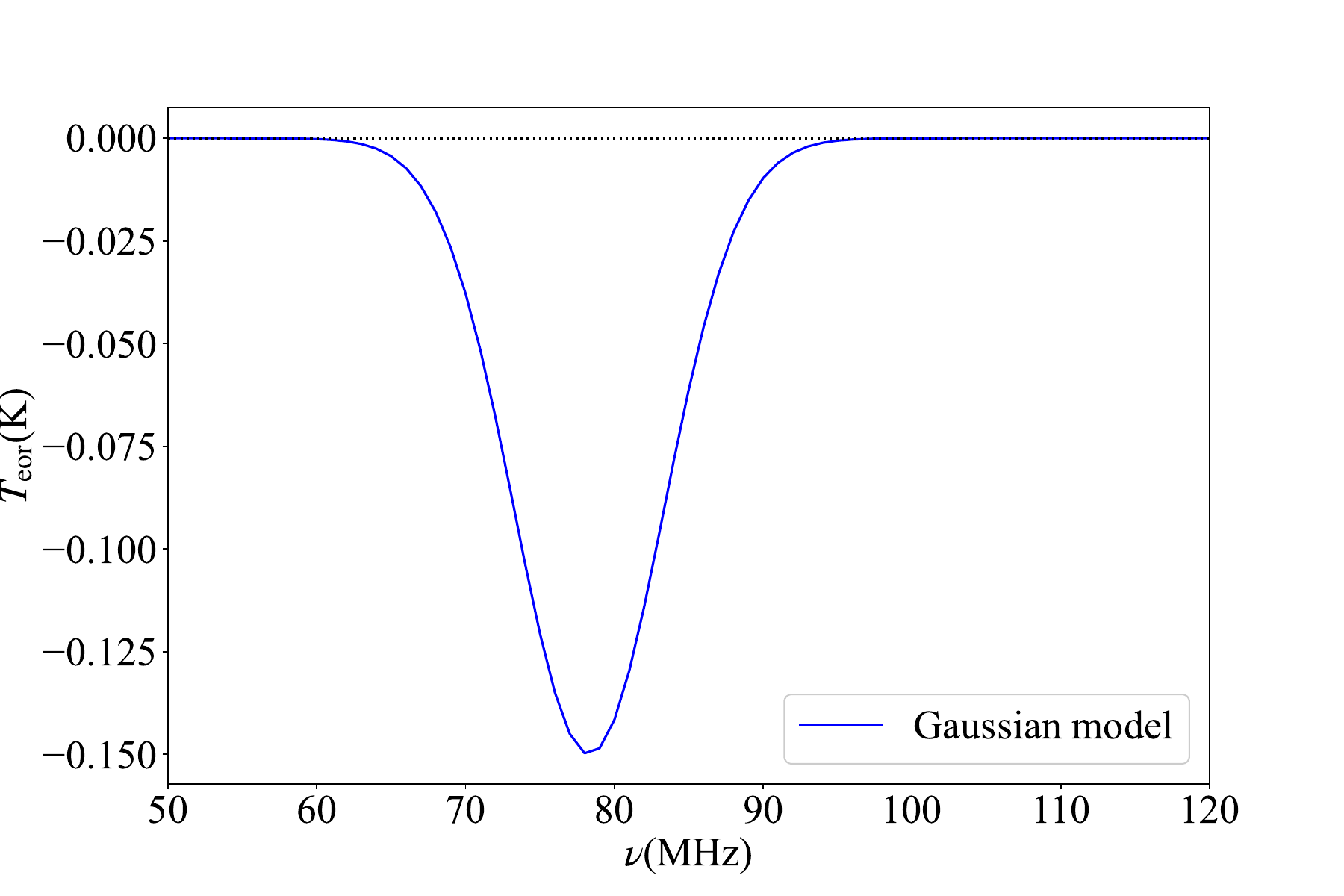}
	\caption{The Gaussian brightness temperature profile of the 21\,cm signal model.}
	\label{fig:Gaussian_model}
\end{figure}

This study aims to detect the absorption features during the CD and EoR, arising from the combined effects of heating by the radiation of first-generation stars and Lyman-alpha coupling \citep[][]{pritchard201221}. This signal is widely perceived as a Gaussian profile and is typically expected within the frequency range of approximately 50–120\,MHz. Consequently, many researchers have employed a simple Gaussian model as the input signal in simulations to evaluate their methods for signal extraction \citep[][]{bernardi2015foreground, presley2015measuring, bernardi2016bayesian, shi2022lunar, kirkham2024bayesian, anstey2021general}. Therefore, we also simulate the cosmic 21\,cm absorption feature using a Gaussian model, described by the following equation:
\begin{equation}
	T_\mathrm{eor}(\nu)=A\exp\left[-\dfrac{(\nu-\nu_c)^2}{2\omega^2}\right],
	\label{eq:Gaussian_model}
\end{equation}
where amplitude $A=-0.150$\,K, centre frequency $\nu_c=78.3$\,MHz, and width $\omega=5$\,MHz \citep[][]{liu2024detecting}. The Gaussian model is shown in Fig.~\ref{fig:Gaussian_model}.

In practice, we also utilized \texttt{GLOBALEMU} to test a wide range of theoretical models \citep[][]{bevins2021globalemu}. This open-source software generates theoretical 21\,cm signal models rapidly by taking seven specific cosmological parameters as input. Our tests on various theoretical models revealed that when the models are close to the Gaussian profile, the VZOP method outperforms common polynomial fitting. However, for models that deviate significantly from the Gaussian profile, neither VZOP nor conventional polynomial fitting proves effective. The primary issue lies in the inadequacy of the chosen fitting model, highlighting the need to explore more appropriate alternatives. For example, we could simultaneously parametrize both the absorption trough at lower frequencies and the emission peak at higher frequencies, similar to the approach taken by \citet{mozdzen2016limits}, who fit the high-frequency emission spectrum with a Gaussian model. Alternatively, theoretical formulae could be directly incorporated into the fitting process \citep[][]{gu2020direct}. Moreover, in many cases, a significant portion of the absorption trough from the EoR falls outside the target frequency range, which may indicate that several experiments were inadequately designed from the outset. In summary, selecting an appropriate fitting model is a complex issue, which we will address in future works.

\subsection{Antennas}
\begin{figure}
	\includegraphics[trim=2cm 14cm 2cm 2cm, clip,width=\columnwidth]{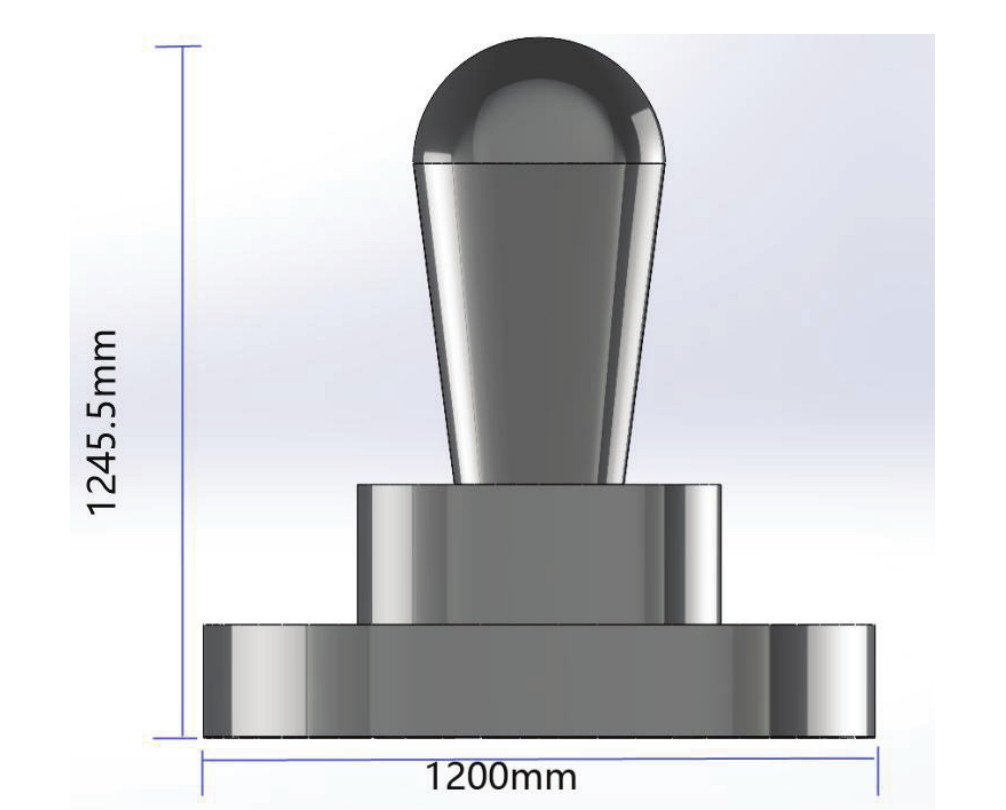}
	\caption{The structure diagram of the ice cream antenna.}
	\label{fig:disc_cone_antenna}
\end{figure}

On the high-frequency daughter satellite of DSL, an ice cream antenna will be employed for global 21\,cm signal observations, as illustrated in Fig.~\ref{fig:disc_cone_antenna}. This antenna features a wide field of view, allowing it to mitigate the impact of small-scale fluctuations on the global signal. Simultaneously, the antenna exhibits minimal chromaticity to avoid introducing additional spectral structures.

\begin{figure}
	\includegraphics[width=\columnwidth]{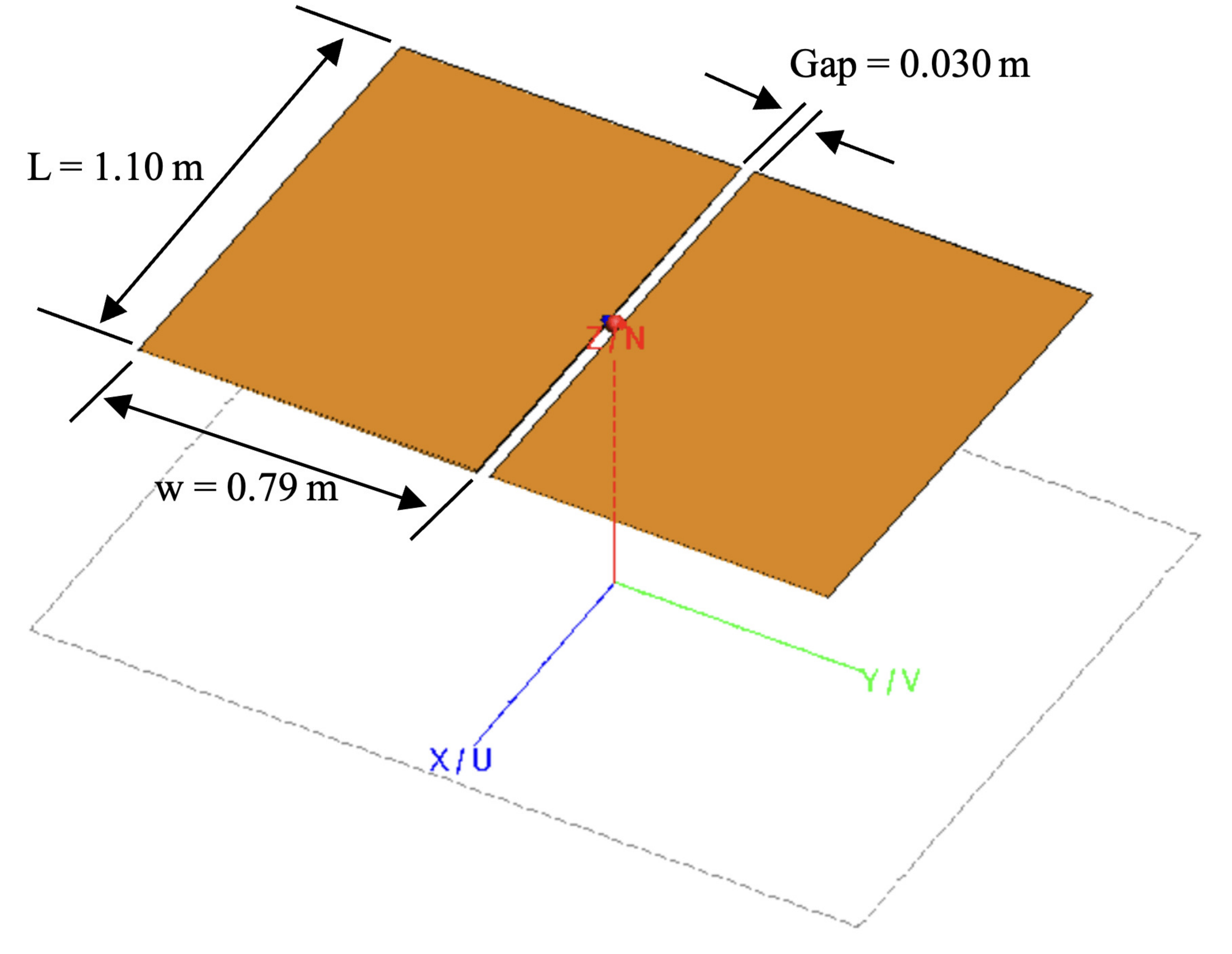}
	\caption{The structure diagram of the blade antenna.}
	\label{fig:blade_antenna}
\end{figure}

\begin{figure*}
	\centering
	\includegraphics[width=\linewidth]{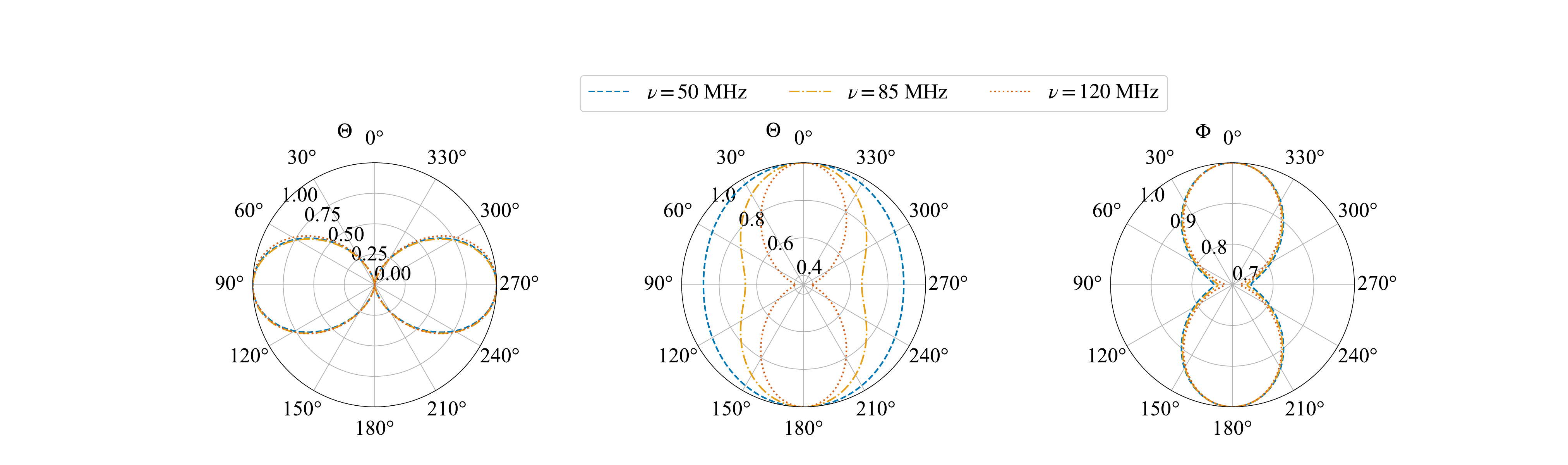}
	\caption{The cross section of the power beam profiles at 50\,MHz (blue dashed), 85\,MHz (orange dash-dotted) and 120\,MHz (vermilion dotted). The left panel shows the gain values of the ice cream antenna at various zenith angles $\Theta$, with values for the range $180^\circ-360^\circ$ indicating the opposite direction of $\Phi$. The middle and right panels show the gain values of the blade antenna at different zenith angles (with fixed azimuth angle $\Phi=0^\circ$ and $\Phi=180^\circ$) and different azimuth angles (with fixed zenith angle $\Theta=30^\circ$), respectively.}
	\label{fig:chromaticity}
\end{figure*}

The ice cream antenna is a rotationally symmetric antenna that naturally lends itself to use in VZOP. However, to demonstrate that asymmetric antennas can also be applied in VZOP, we additionally tested a blade antenna, depicted schematically in Fig.~\ref{fig:blade_antenna}. In Fig.~\ref{fig:chromaticity}, we present cross-sections of the power beam patterns for two antennas. The left panel illustrates that the ice cream antenna employed by DSL exhibits favourable frequency independence. The blade antenna, assuming a ground plane as an infinite Perfect Electric Conductor, also shows good frequency independence when on the ground, with its gain approaching zero near the horizon. However, this effect diminishes in space, leading to noticeable frequency dependence.

\subsection{Thermal noise}
\begin{figure}
    \centering
	\includegraphics[width=\columnwidth]{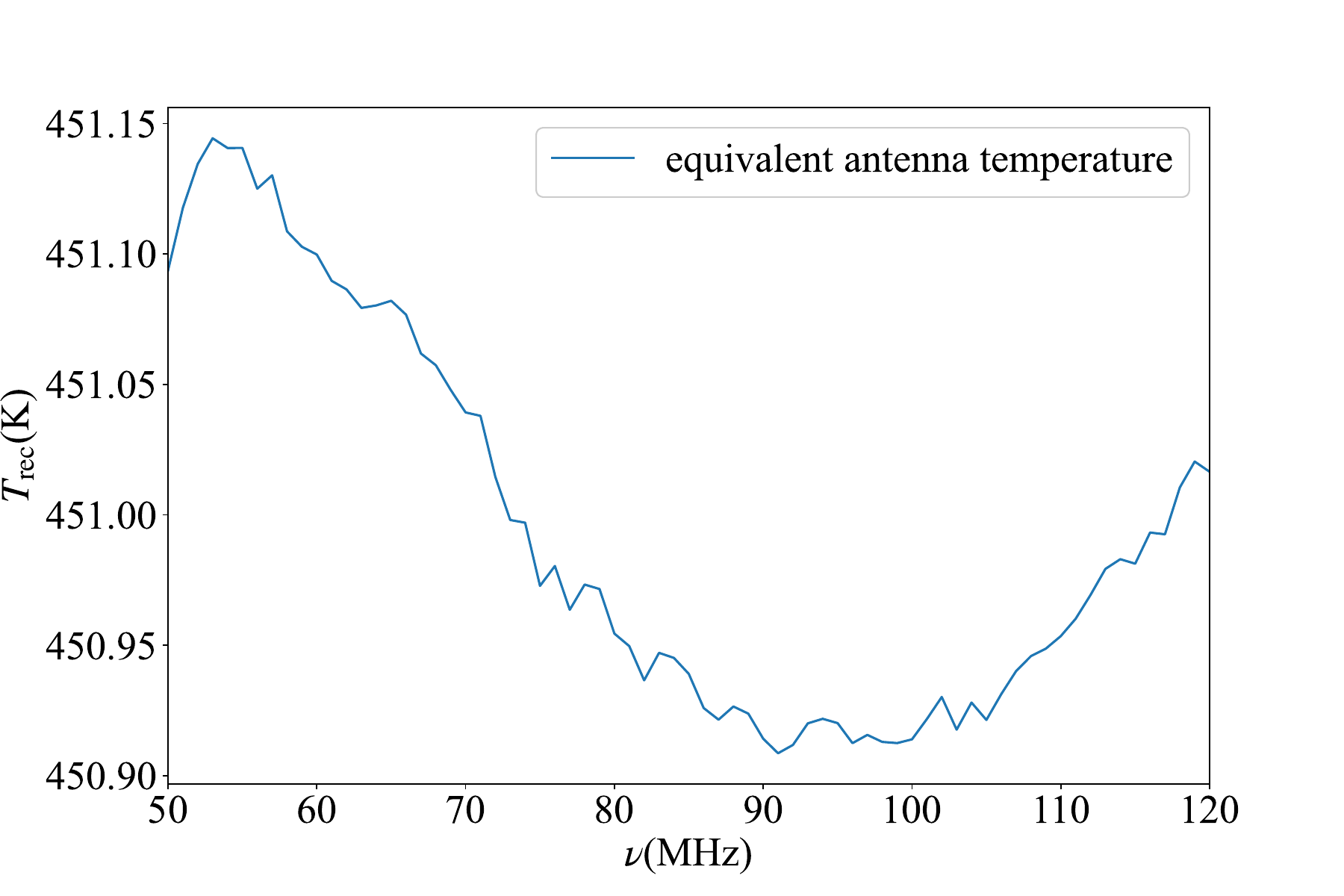}
	\caption{Equivalent antenna temperature caused by impedance mismatch on the high-frequency antenna of the DSL.}
	\label{fig:equivalent_antenna_temperature}
\end{figure}

The system always exhibits thermal noise, and this noise has a minimum value. For a given system component, there is no better system whose RMS (Root Mean Square) fluctuation of systematic errors is less than this minimum value. The minimum value of thermal noise is
\begin{equation}
	\sigma_n(\nu)=\dfrac{T_\mathrm{sys}(\nu)}{\sqrt{N_\mathrm{m}\Delta\nu t_\mathrm{int}}}
	\label{eq:thermal noise},
\end{equation}
Where the $T_\mathrm{sys}(\nu)$ is the system temperature, $N_\mathrm{m}=1$ is the number of independent measurements, $\Delta\nu$ is the frequency channel bandwidth and $t_\mathrm{int}$ is the integration time. In this work we set $\Delta\nu=1\,\mathrm{MHz}$ and $t_\mathrm{int}=10\,\mathrm{d}$. The system temperature consists of two components, that is
\begin{equation}
	T_\mathrm{sys}(\nu)=T_\mathrm{sky}(\nu)+T_\mathrm{rec}(\nu)
	\label{eq:system temperature},
\end{equation}
Where $T_\mathrm{sky}(\nu)$ is the integrated sky temperature and $T_\mathrm{rec}(\nu)$ is the equivalent antenna temperature caused by the noise in the receiver. The input and output impedance mismatch results in reflection, denoted by the reflection coefficient $\Gamma(\nu)$. The relationship between $T_\mathrm{rec}(\nu)$ and the reflection coefficient $\Gamma(\nu)$ is given by
\begin{equation}
	T_\mathrm{rec}(\nu)=\dfrac{T^0_\mathrm{rec}}{1-\lvert\Gamma(\nu)\rvert^2}
	\label{eq:equivalent_antenna_temperature},
\end{equation}
where $T^0_\mathrm{rec}$ represents the receiver noise temperature, which for the current design of DSL is $T^0_\mathrm{rec}\approx{450\,\mathrm{K}}$. The values of the reflection coefficient $\Gamma$ are obtained from simulations, and its frequency-dependent relationship is depicted in Fig.~\ref{fig:equivalent_antenna_temperature}.

In the following discussion, we assume that $T_\mathrm{rec}(\nu)$ has already been subtracted and only contributes to the thermal noise in the antenna temperature. The specific calibration strategy is an important and complex issue, with some experiments having proposed their approaches \citep[][]{roque2021bayesian, de2022reach, razavi2023receiver, bull2024rhino}. However, these details lie beyond the scope of this paper.

\subsection{Simulation process}
\label{subsec:simulation_process}
The satellite takes approximately 8246 seconds to complete one orbit around the Moon. On average, there is a period of roughly 10\% of this time when both the Earth and the Sun are simultaneously shielded by the Moon. This interval is considered the optimal observation period and is referred to as the "doubly good time" \citep[][]{shi2022lunar}. Consequently, we set the antenna's rotation period to 80 seconds, allowing for an 800-second observation during each Lunar orbit. To achieve a cumulative observation time of $8.64\times10^{5}$ seconds (i.e., 10 days), the satellite must complete 1080 orbits around the Moon, which takes approximately 103 days.

Although the antenna observations during the "doubly good time" are not affected by radiation from the Sun and Earth RFI, there is still radiation from the Moon. If we take into account the Lunar radiation, the simulated sky temperature can be expressed as:
\begin{equation}
	T_\mathrm{sky}(\theta,\phi)=\mathcal{S}(\theta)T_b(\theta,\phi)+\left[1-\mathcal{S}(\theta)\right]T_\mathrm{Lunar}(\theta,\phi),
	\label{eq:cosider Lunar radiation}
\end{equation}
where $T_\mathrm{Lunar}(\nu,\theta,\phi,t)$ is the Lunar radiation temperature, which consists of two parts
\begin{equation}
	T_\mathrm{Lunar}(\theta,\phi)=T_\mathrm{Moon}+T_\mathrm{refl}(\theta,\phi),
	\label{eq:Lunar radiation}
\end{equation}
where $T_\mathrm{Moon}$ is the Lunar intrinsic temperature and $T_\mathrm{refl}(\theta,\phi)$ is the reflected sky temperature of the Lunar surface. 

\citet{mckinley2018measuring} assumed that the Lunar intrinsic temperature remains constant within the 72–230\,MHz range, citing its decreasing frequency dependence at lower frequencies \citep[][]{krotikov1964radio}, which becomes negligible at meter wavelengths \citep[][]{baldwin1961thermal}. Using the Murchison Widefield Array (MWA) and the Lunar occultation method to measure the global 21\,cm signal, they determined the Lunar intrinsic temperature to be $180 \pm 12$\,K. \citet{tiwari2023measuring} further refined the constraints on the Lunar intrinsic temperature, obtaining two results using different methods: $184.4 \pm 2.6$\,K and $173.8 \pm 2.5$\,K. For simplicity, we adopt an intrinsic Lunar temperature of $T_\mathrm{Moon}=180\,\mathrm{K}$. The reflected sky temperature from the Lunar surface is more complex and depends on the topography and composition of the Lunar soil. \citet{le2023lunar} used data provided by the Lunar Orbiter Laser Altimeter (LOLA) to create a topographic map of the Lunar far-side \citep[][]{smith2010lunar}. However, there is currently no reflectance data for the Lunar far-side in the CD-EoR frequency band. Nevertheless, reflectance coefficient data for the Lunar near-side has been measured through radar reflection \citep[][]{evans1969radar}, which we can temporarily use as a substitute. Research indicates that the Lunar near-side surface in the CD-EoR frequency band can be treated as a mirror-like reflection with a reflectance of about 7\%. Therefore, in this study, we assume that the Lunar far side is also a mirror-like reflection with a constant reflectance of 7\%. Additionally, we do not consider variations in surface topography but instead assume it to be a smooth spherical surface. We calculate the Lunar reflected brightness temperature $T_\mathrm{refl}(\theta,\phi)$, and the detailed derivation process is provided in the \ref{sec:LR}. The assumed Lunar reflection model here is a simplified one. We will discuss the impact of Lunar reflection in detail in future work.

\section{Results}\label{sec:Results}
\subsection{No Lunar radiation}
\begin{figure*}
    \centering
	\includegraphics[width=\linewidth]{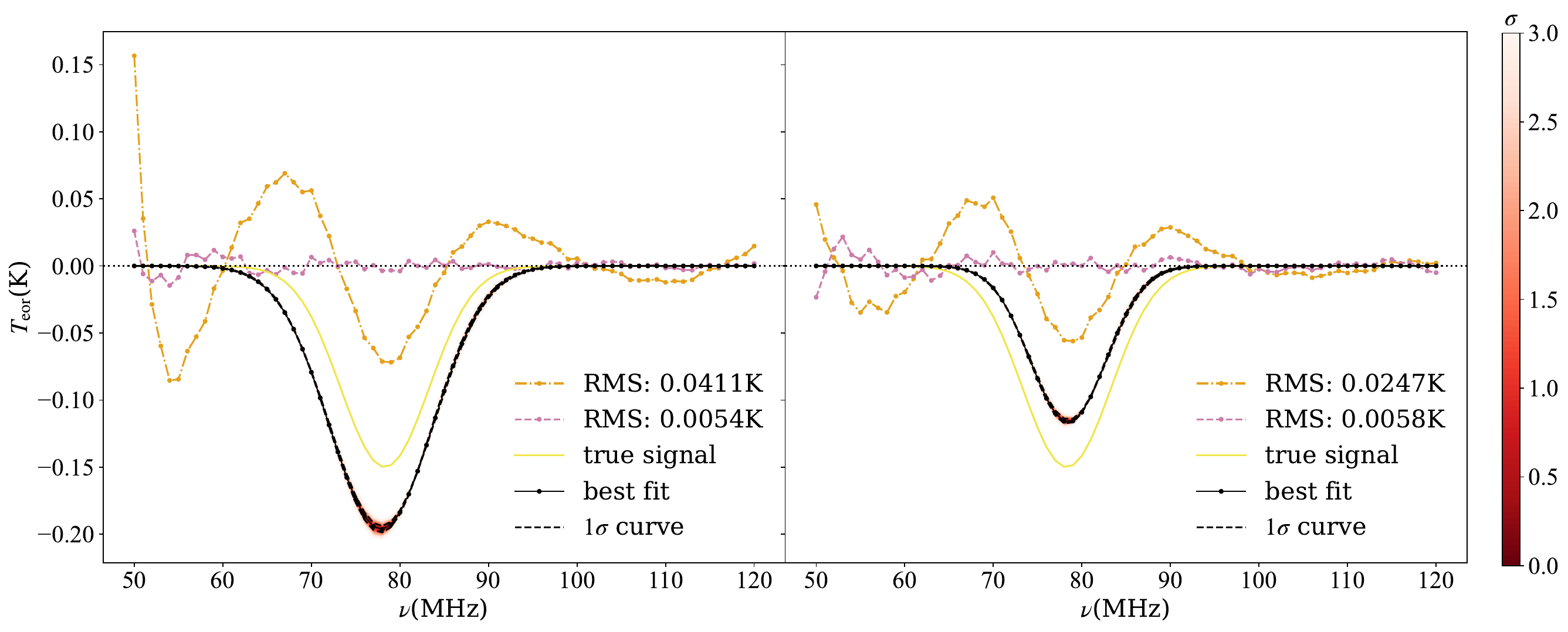}
	\caption{The results of common polynomial fitting based on the ice cream antenna (left) and the blade antenna (right). The orange dash-dotted line represents the residuals obtained when fitting and subtracting only the foreground, while the reddish-purple dashed line shows the residuals when fitting and subtracting both the foreground and the signal. The yellow solid line represents the input signal, and the black solid line is the best-fit line. Additionally, multiple lines representing different levels of errors are depicted, with varying shades of colour to indicate error magnitude, as detailed in the colour bar to the right of the figure. For clarity, two black dashed lines specifically denote the 1$\sigma$ error lines.}
	\label{fig:polynomial_recovered}
\end{figure*}

Firstly, we neglect the Lunar radiation and obtain the simulated antenna temperature data. For comparison, we initially present the fitting results obtained using the common polynomial fitting in Fig.~\ref{fig:polynomial_recovered}. When fitting and subtracting only the foreground, the fluctuations in the residuals reach several tens of millikelvin. Considering an integration period of 10 days, the thermal noise contributes at most a few millikelvins. Thus, we have reason to suspect the presence of structure in the antenna temperature. Therefore, we include the 21\,cm signal model and refit, resulting in significantly reduced residuals. However, there remains a bias between the sampled signal and the input signal. The fitting results for the ice cream antenna (left panel) exhibit a deeper feature than the input signal, while the blade antenna (right panel) exhibits a shallower feature compared to the input signal. Regardless of the antenna used, the common polynomial fitting algorithm fails to eliminate the additional structure introduced due to chromaticity, which is the same as our previous work \citep[][]{liu2024detecting}.

\begin{figure*}
    \centering
	\includegraphics[width=\linewidth]{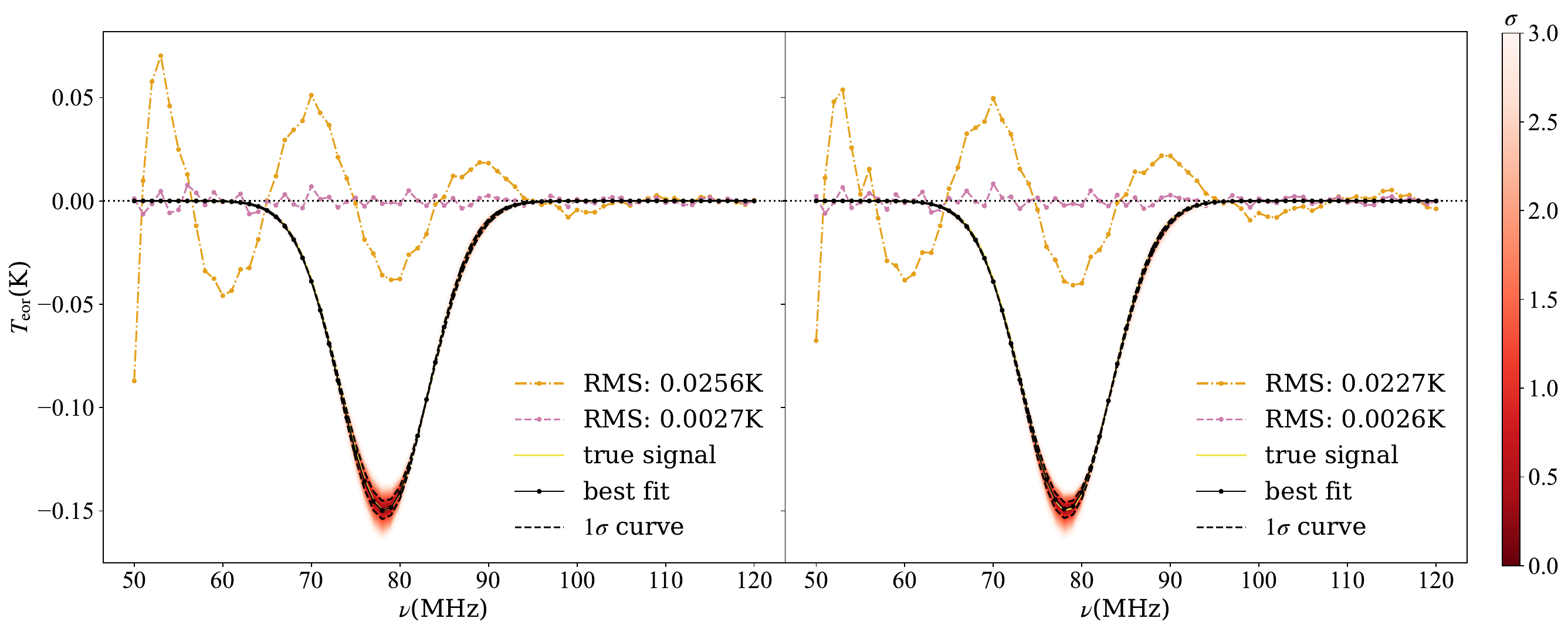}
	\caption{The fitting results of VZOP with 10 declination bins based on the ice cream antenna (left) and the blade antenna (right). The orange dash-dotted line represents the residuals obtained when fitting and subtracting only the foreground, while the reddish-purple dashed line shows the residuals when fitting and subtracting both the foreground and the signal. The yellow solid line represents the input signal, and the black solid line is the best-fit line. Additionally, multiple lines representing different levels of errors are depicted, with varying shades of colour to indicate error magnitude, as detailed in the colour bar to the right of the figure. For clarity, two black dashed lines specifically denote the 1$\sigma$ error lines.}
	\label{fig:VZOP_recovered}
\end{figure*}

We utilized VZOP with 10 declination bins to fit the antenna temperatures, and the outcomes are depicted in Fig.~\ref{fig:VZOP_recovered}. Regardless of the antenna used, the 21\,cm signal recovered with VZOP closely matches the input signal. This indicates that VZOP has effectively incorporated sufficient beam-related information into the model, successfully removing the additional structure introduced by imperfect antennas. Furthermore, the two residual lines in the left panel match those in the right panel, underscoring the effectiveness of VZOP in mitigating the antenna's chromaticity effect, even for the asymmetric antenna. In contrast, the residual lines in the left and right panels of Fig.~\ref{fig:polynomial_recovered} are different.

We conducted additional tests to demonstrate the robustness of VZOP and its advantages over common polynomial fitting. To maintain focus in the main text, these results are provided in \ref{sec:robust}.

\subsection{Considering the Lunar radiation}
\begin{figure}
	\includegraphics[width=\columnwidth]{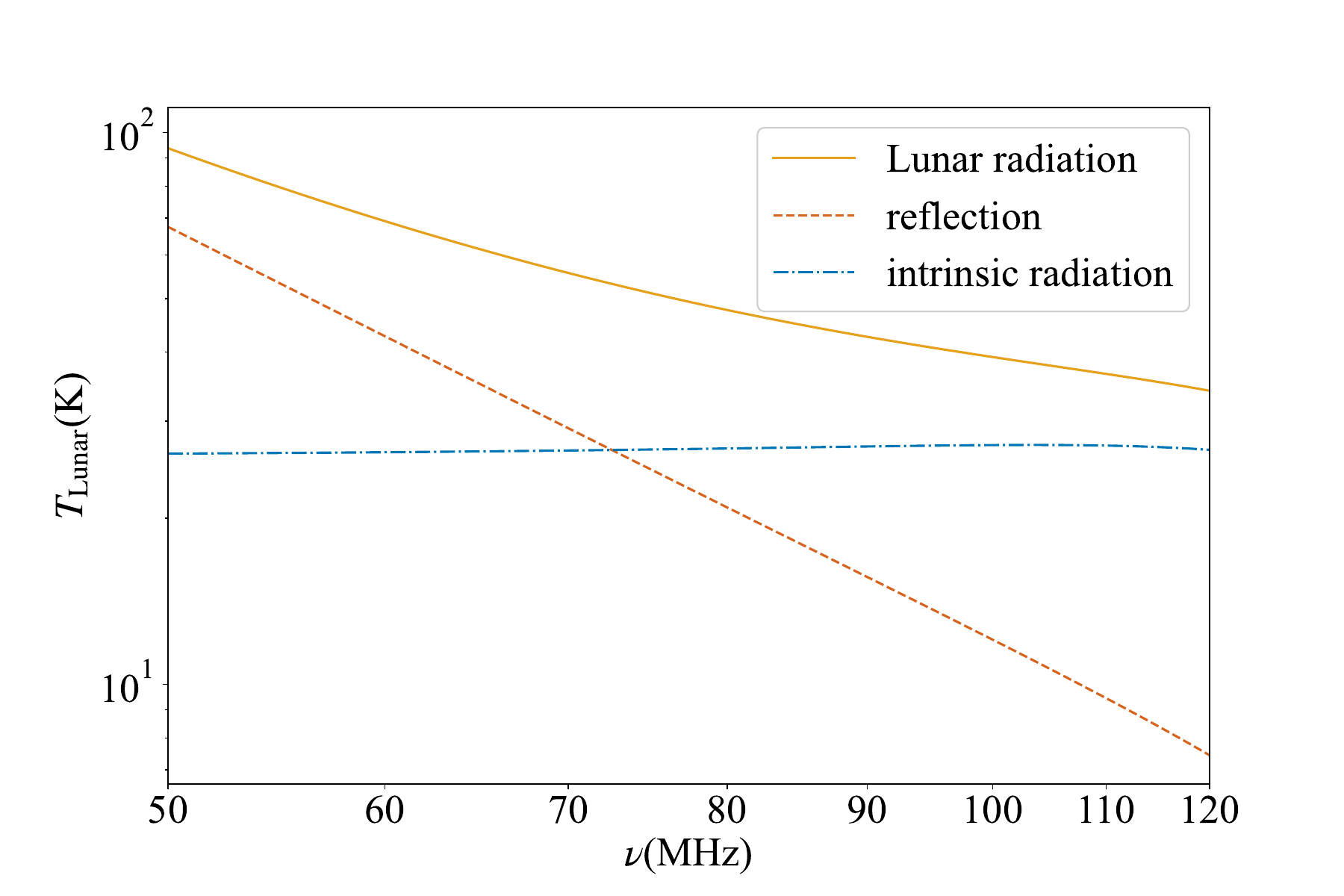}
	\caption{The contribution of Lunar radiation to the antenna temperature using an ice cream antenna. The solid line represents the contribution of Lunar radiation to the antenna temperature, the dashed line indicates the contribution from the Moon's reflection to the sky temperature, and the dash-dotted line represents the contribution from the Moon's intrinsic blackbody radiation. The reflected temperature is essentially a smooth power-law spectrum, but the intrinsic radiation is an approximately horizontal line, demonstrating that Lunar radiation does not follow a power-law spectrum.}
	\label{fig:Lunar_radiation}
\end{figure}

\begin{figure}
	\includegraphics[width=\columnwidth]{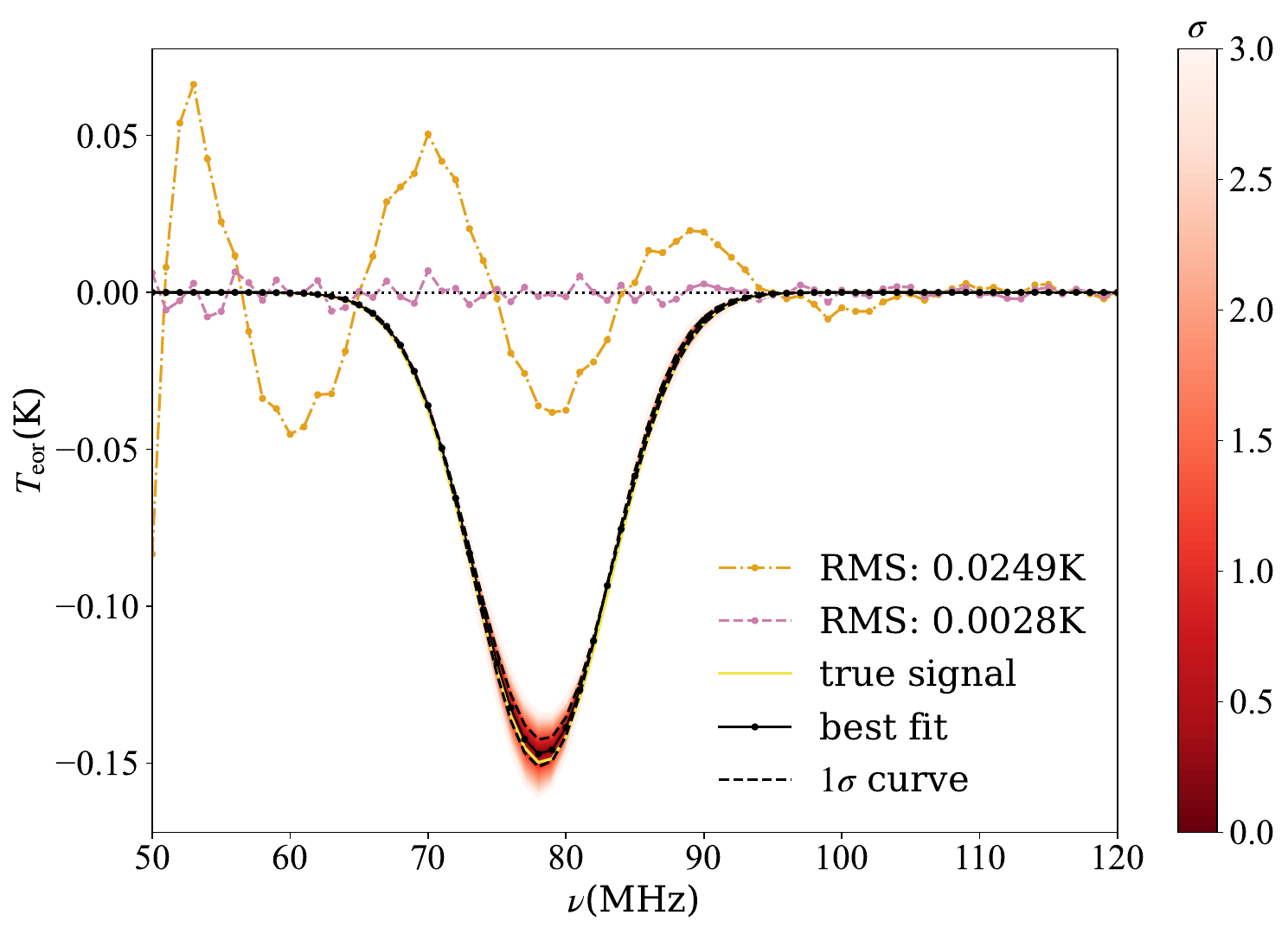}
	\caption{The fitting results of VZOP with 10 declination bins based on the ice cream antenna considering the Lunar radiation. The orange dash-dotted line represents the residuals obtained when fitting and subtracting only the foreground, while the reddish-purple dashed line shows the residuals when fitting and subtracting both the foreground and the signal. The yellow solid line represents the input signal, and the black solid line is the best-fit line. Additionally, multiple lines representing different levels of errors are depicted, with varying shades of colour to indicate error magnitude, as detailed in the colour bar to the right of the figure. For clarity, two black dashed lines specifically denote the 1$\sigma$ error lines.}
	\label{fig:fit_with_albedo}
\end{figure}

Fig.~\ref{fig:Lunar_radiation} illustrates the contribution of Lunar radiation to the antenna temperature when observed using an ice cream antenna. The reflection of the sky temperature by the moon still manifests as a power-law spectrum. Due to the antenna's chromaticity, the intrinsic radiation from the Moon is only an approximate horizontal line, not precisely a constant. Because of this reason, Lunar radiation does not exhibit a power-law spectrum. However, as shown in Fig.~\ref{fig:fit_with_albedo}, VZOP can still recover the 21\,cm signal. Moreover, the two residual lines closely match the two residuals in the left panel of Fig.~\ref{fig:VZOP_recovered}, indicating that Lunar radiation has not affected VZOP.

\section{Non-Ideal Situations}\label{sec:NIS}
The above results represent ideal scenarios; however, real-world situations may be more complex. To validate the robustness of our algorithm, this section considers some non-ideal situations. The antenna temperature models used in this section all account for Lunar radiation.

\subsection{Inaccurate antenna beam pattern}\label{subsec:IABP}
The preceding analyses have all assumed that the antenna beam is precisely known, which is practically impossible. Now, we assume that there are errors in the antenna model used in VZOP while employing a precise antenna model during observation. Since antenna errors are systematic, although they are calibrated for, there will still be some residual errors, which are not entirely random. The ice cream antenna is symmetrical, the simulated antenna beam data are along the zenith angle direction, with a data interval of $1^\circ$. We assume that there are entirely random relative errors at different zenith angles $\theta$ but follow a cosine function at different frequencies $\nu$
\begin{equation}
	\mathrm{e}_B(\nu,\theta)=\cos{\left(\frac{2\pi}{\nu_p}\nu\right)}\times\mathrm{e}_B(\nu_0,\theta),
	\label{eq:relative_error}
\end{equation}
where $\nu_0=50\,\mathrm{MHz}$ and period $\nu_p=10\,\mathrm{MHz}$. We generate random relative errors at different zenith angles around the frequency $\nu_0$ and then construct the entire error model according to equation~(\ref{eq:relative_error}).

\begin{figure}
	\includegraphics[width=\columnwidth]{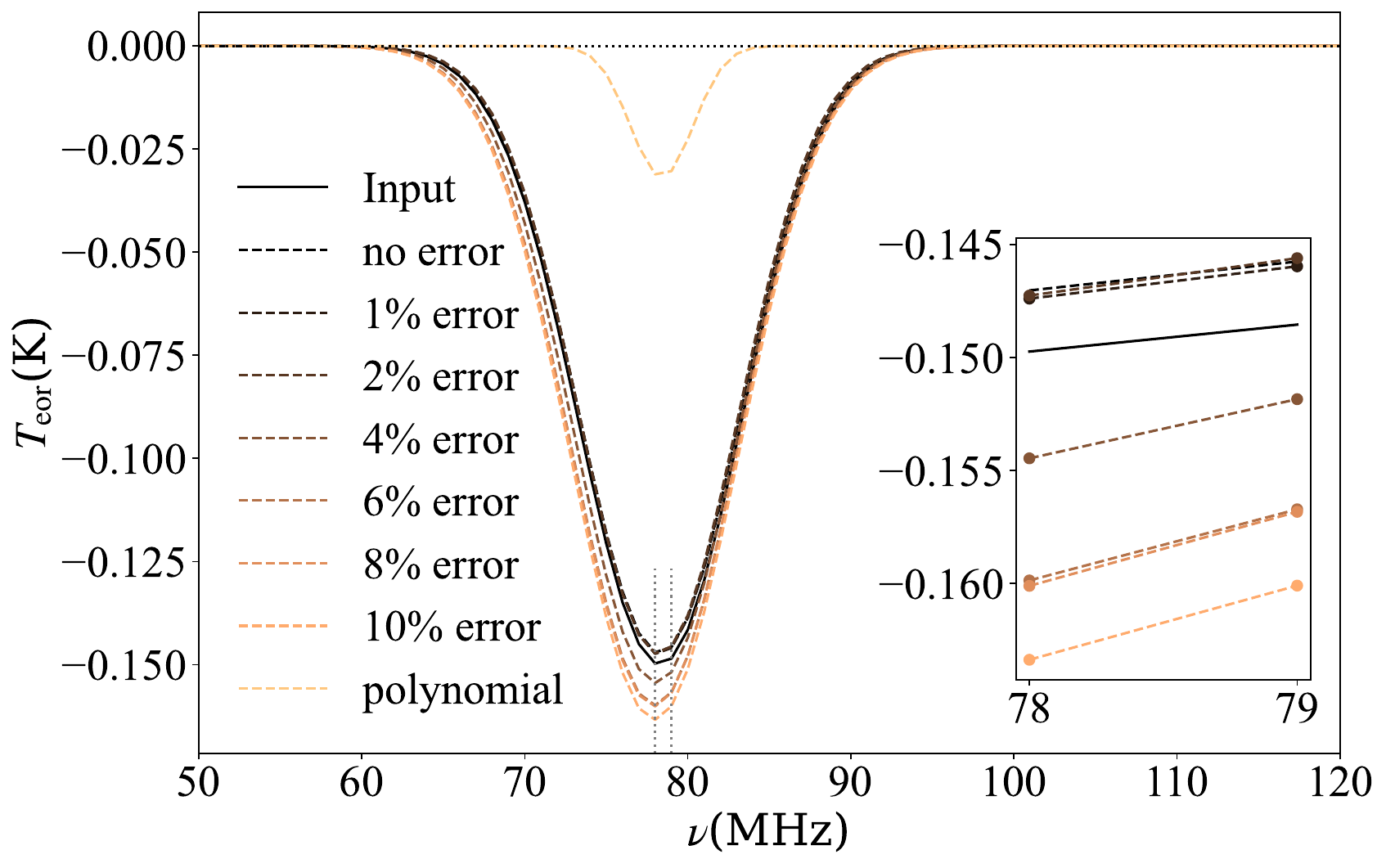}
	\caption{The fitting results of the VZOP method using 10 declination bins in the presence of beam errors for the ice cream antenna. The black solid line represents the input signal model, while the dashed lines represent the fitting results under different levels of errors, with colours ranging from dark to light indicating increasing errors. For comparison, we also included the fitting results of the common polynomial.}
	\label{fig:antenna_errors}
\end{figure}

Fig.~\ref{fig:antenna_errors} illustrates the fitting results under different error scenarios, alongside the result obtained using the common polynomial fitting for comparison. It is worth noting that since the mean beam within each declination bin is essentially an average over all gain points within that bin, the actual relative error of the mean beam is smaller than the error at each zenith angle. As the error magnitude increases, there is a slight decrease in the accuracy of signal recovery. However, even with a 10\% error, the fitting results remain highly accurate. Compared to the left panel in Fig.~\ref{fig:polynomial_recovered}, the result obtained by the common polynomial fitting here is only $-0.03\,K$, which completely fails to identify the 21\,cm signal after considering Lunar radiation. This may be attributed to the distortion of foreground power-law characteristics by Lunar intrinsic blackbody radiation, while VZOP, possessing greater degrees of freedom, is hardly affected. This underscores the superiority of VZOP.

\subsection{RFI contamination}
One of the key objectives of launching radio astronomy receivers into Lunar orbit is to avoid interference from terrestrial RFI, although this does not guarantee complete avoidance of RFI. On one hand, the antenna and receiver systems themselves may introduce RFI. On the other hand, if during the observation time, Earth is not entirely on the other side of the Moon, it may result in RFI leakage from Earth. To verify the robustness of VZOP, we assume that the antenna temperature includes the influence of RFI and re-extract the 21\,cm signal.

\begin{figure*}
    \centering
	\includegraphics[width=\linewidth]{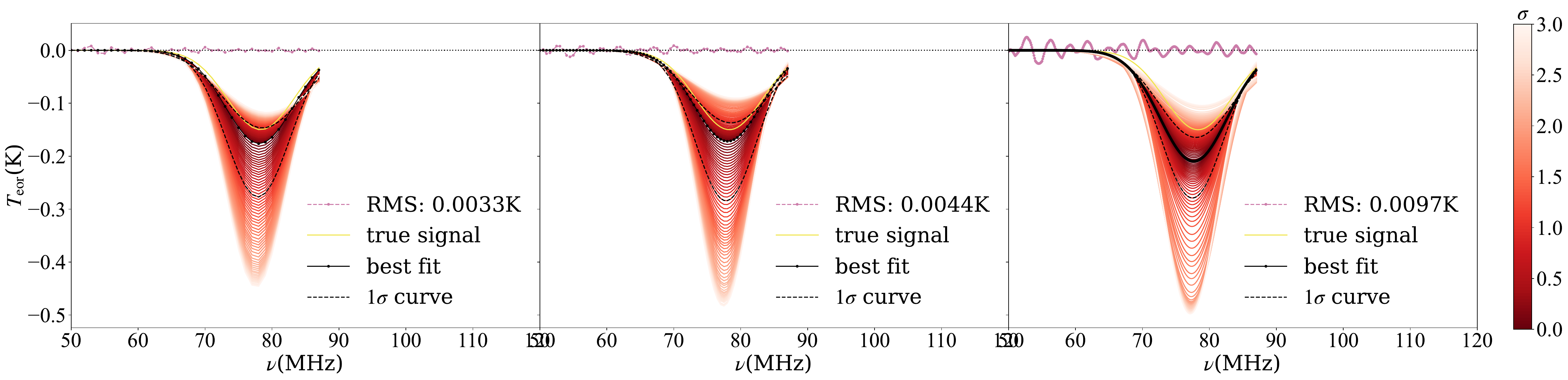}
	\caption{The fitting results of VZOP with 10 declination bins based on the ice cream antenna with frequency resolutions of 1\,MHz (left), 0.5\,MHz (middle), and 0.1\,MHz (right). The reddish-purple dashed line shows the residuals when fitting and subtracting both the foreground and the signal. The yellow solid line represents the input signal, and the black solid line is the best-fit line. Additionally, multiple lines representing different levels of errors are depicted, with varying shades of colour to indicate error magnitude, as detailed in the colour bar to the right of the figure. For clarity, two black dashed lines specifically denote the 1$\sigma$ error lines. The fitting results at different frequency resolutions show no significant difference.}
	\label{fig:removeFM_VZOP_recovered}
\end{figure*}

Within our target frequency range, one of the most significant RFI sources is FM radio broadcasting, which operates within the frequency range of 88-108\,MHz \citep[][]{offringa2013lofar, offringa2015low, huang2016radio, sokolowski2016statistics}. FM radio signals are typically strong and easily identifiable. Once the strong RFI is detected, the common practice is to flag it and then remove it for further analysis. We assume that FM radio has leaked into the received antenna temperature data, and then remove all data from 88-120\,MHz. We then apply VZOP to fit the antenna temperature data again. The results, as shown in the left panel of Fig.~\ref{fig:removeFM_VZOP_recovered}, indicate that due to the removal of data containing some 21\,cm signal, the remaining profile is incomplete. Consequently, the accuracy of the re-fitted 21\,cm signal decreases significantly, while the uncertainty increases substantially. This suggests that if the frequency band contaminated by leaked FM radio is removed, and if there indeed exists a 21\,cm signal profile within that band, the fitting performance deteriorates.

\begin{figure}
	\includegraphics[width=\columnwidth]{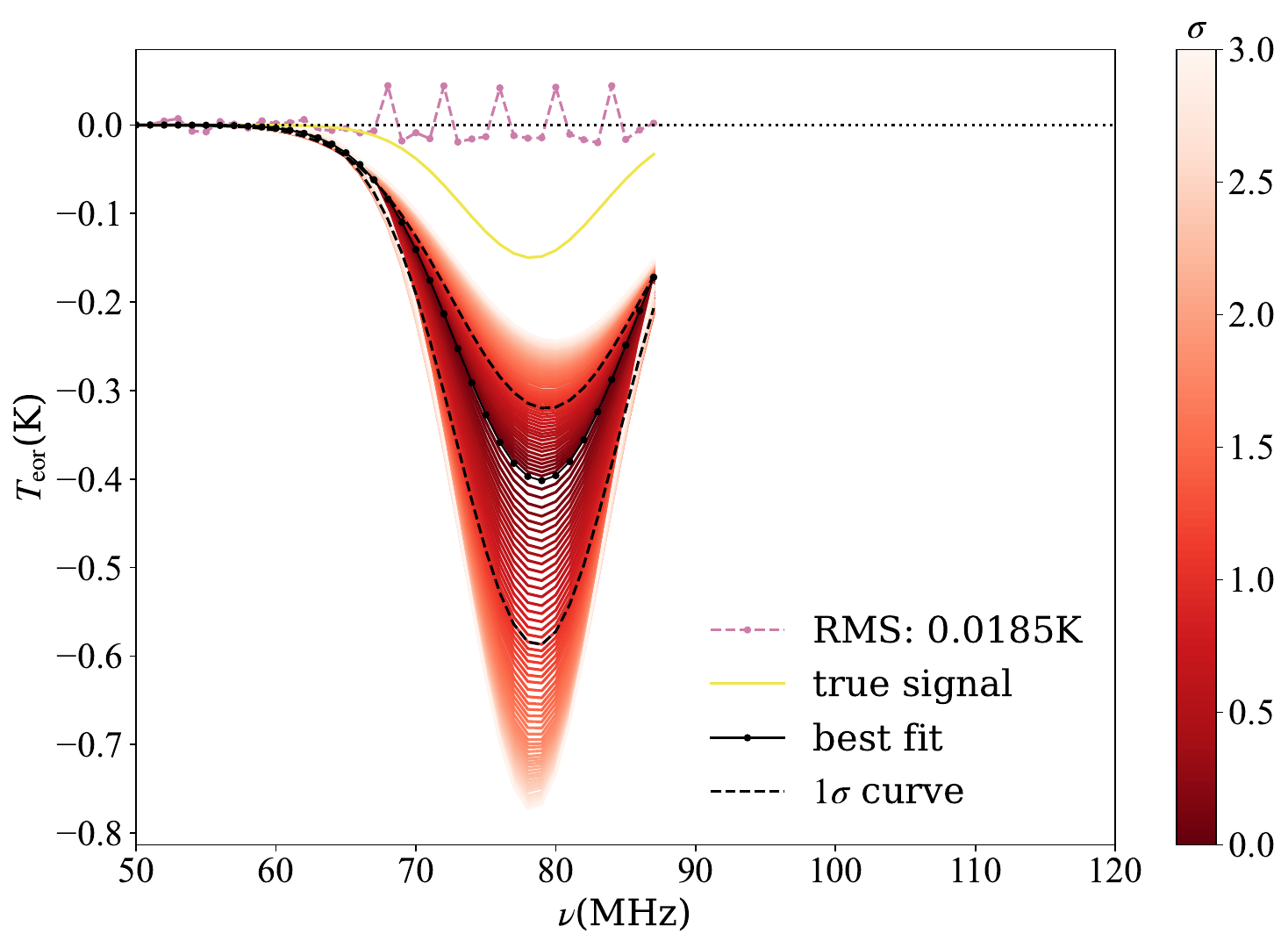}
	\caption{The fitting results of VZOP (10 declination bins) based on the ice cream antenna assuming the presence of 0.05\,K RFIs at 68, 72, 76, 80 and 84\,MHz are not identified (88-120\,MHz has been removed). The reddish-purple dashed line shows the residuals when fitting and subtracting both the foreground and the signal. The yellow solid line represents the input signal, and the black solid line is the best-fit line. Additionally, multiple lines representing different levels of errors are depicted, with varying shades of colour to indicate error magnitude, as detailed in the colour bar to the right of the figure. For clarity, two black dashed lines specifically denote the 1$\sigma$ error lines.}
	\label{fig:fix_weak_rfi}
\end{figure}

\begin{figure}
	\includegraphics[width=\columnwidth]{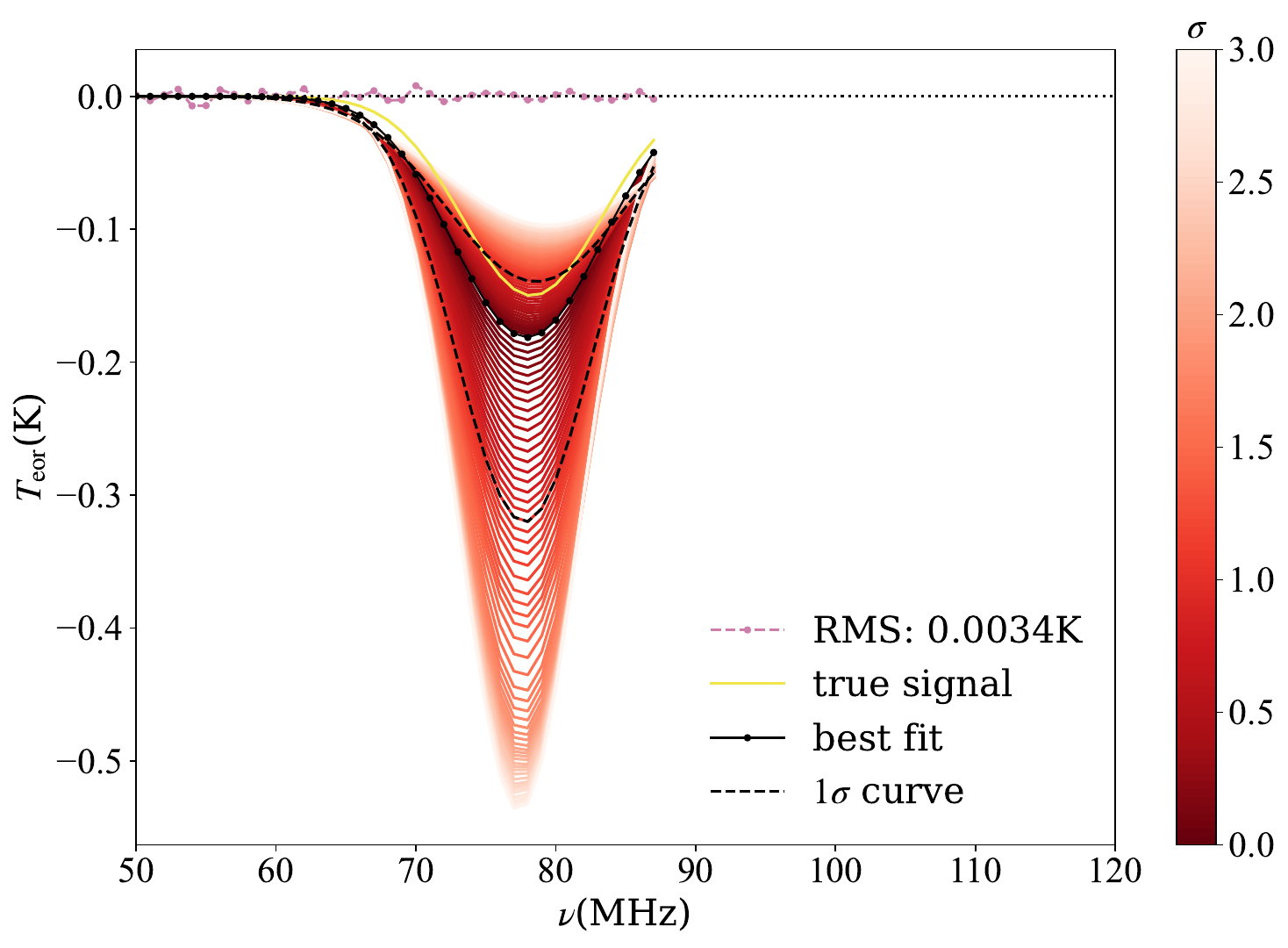}
	\caption{Results of re-fitting using VZOP (with 10 declination bins) after interpolation following the removal of faint RFI in Fig.~\ref{fig:fix_weak_rfi}. The reddish-purple dashed line shows the residuals when fitting and subtracting both the foreground and the signal. The yellow solid line represents the input signal, and the black solid line is the best-fit line. Additionally, multiple lines representing different levels of errors are depicted, with varying shades of colour to indicate error magnitude, as detailed in the colour bar to the right of the figure. For clarity, two black dashed lines specifically denote the 1$\sigma$ error lines.}
	\label{fig:refix_after_remove_rfi}
\end{figure}

Furthermore, to compare the effectiveness of VZOP at different frequency resolutions, we also present the fitting results with frequency resolutions of 0.5\,MHz and 0.1\,MHz in the middle and right panels of Fig.~\ref{fig:removeFM_VZOP_recovered}, respectively. We find that the different frequency resolutions do not significantly affect the results, indicating that we can employ relatively wide frequency resolutions for signal extraction. This approach is advantageous because, besides broad-band RFI like FM radio, there are also many potential narrow-band RFIs that approximate line spectra, such as those generated by instruments themselves. These narrow-band RFIs typically have widths much smaller than 1\,MHz, and fitting with a relatively wide frequency resolution can treat them as line spectra, making processing much easier. If line spectra RFIs are strong, we can still identify and flag them; however, if they are weak, they may go unnoticed. We assume the existence of weak RFIs within the 21\,cm signal band that were not identified and try fitting using VZOP, as shown in Fig.~\ref{fig:fix_weak_rfi}. If there are undetected weak RFIs, they may prevent the extraction of the 21\,cm signal. However, we can also observe that if any RFI exceeds the amplitude of random thermal noise significantly, we can identify them from the residual curve. Therefore, we can now identify and discard data points contaminated by RFIs. Then, we can interpolate the data from the nearest two frequencies and re-fit using VZOP, as shown in Fig.~\ref{fig:refix_after_remove_rfi}. The results closely approximate those in the left panel of Fig.~\ref{fig:removeFM_VZOP_recovered}, indicating the successful removal of the influence of weak RFIs. Therefore, in the case of discrete line-like spectrum RFI, if it is strong, contaminated data points can be replaced by interpolation and then fitted using VZOP. Even if the RFI magnitude is too small to be directly identified, as long as it exceeds random thermal noise, it can be fitted using VZOP first, and then indirectly identified based on the residual curve.

\section{Discussions and Conclusions}\label{sec:DAC}
DSL is a project that aims to launch a constellation of satellites in Lunar orbit, the high-frequency daughter satellite of which will detect the cosmic global 21\,cm signal. When the satellite is positioned on the far side of the Moon, and the Sun is simultaneously obscured by the Moon, it provides the optimal observation window, referred to as the "doubly good time." We attempt to apply VZOP, an optimized polynomial fitting algorithm designed to remove additional structures caused by the frequency dependence of the antenna beam, to fit the simulated observation data of DSL.

The ice cream antenna that will be carried by the satellite is a symmetric antenna, naturally suitable for using VZOP. Asymmetric antennas can also use VZOP, provided that the satellite has a fast rotation in theory. However, a considerable body of research results suggest that non-rotating antennas have little to no impact on signal extraction by VZOP (refer to \ref{sec:SR} for more details). Therefore, current findings indicate that satellite rotation may not be necessary.

We assume observations only during "doubly good time", with a total integration time of 10 days. Initially, we obtain the simulated antenna temperature data assuming no radiation from the Moon. Common polynomial fitting fails to accurately extract the 21\,cm signal due to the frequency dependence of the antenna beam. As anticipated, when using the VZOP algorithm, both the ice cream and asymmetric blade antennas can precisely recover the signal. The VZOP algorithm outperforms common polynomial fitting.

In fact, the Moon also emits radiation, which includes intrinsic blackbody radiation and reflection of sky temperature. The Lunar radiation causes the antenna temperature to deviate from the original power-law spectrum. Nevertheless, we successfully recovered the 21\,cm signal using VZOP, and the two residual curves, with and without consideration of Lunar radiation, were very similar. This suggests that Lunar radiation did not affect the signal recovery process of VZOP.

The above results are derived from ideal conditions. Using a model that includes Lunar radiation as the basic model, we also examined some non-ideal scenarios. Assuming that the relative error of the antenna beam is completely random at different zenith angles but follows a cosine model at different frequencies, we can provide fitting results of VZOP under various error conditions. Even with a relative error amplitude of 10\%, VZOP can still provide highly accurate fitting results, indicating its capability to recover signals without precise beam data.

If the observation timing is not strictly controlled within the "doubly good time" window, leading to Earth's RFI leaking into the antenna, it may significantly impact signal extraction. Assuming leakage of FM radio from Earth, we remove the data in the 88-120\,MHz range and directly fit the remaining data. The results show decreased accuracy and increased uncertainty. This result indicates that if a significant amount of FM radiation leaks into the receiver, the fitting results will be severely affected. Therefore, it is imperative to ensure that the Earth is fully shielded by the Moon. Control over observation time should be stricter, with observations ceasing earlier when the Earth is about to come into view.

There is no significant difference in fitting results with different frequency resolutions, allowing us to choose a wider frequency resolution for easier handling of line-shaped spectrum RFI. Strong line-spectrum RFI can be readily identified and removed, followed by interpolation, and finally fitted using VZOP. Weaker line-spectrum RFI cannot be directly identified but can be recognized through the residual curves obtained after fitting.

In conclusion, the serene radio environment on the Lunar far side presents a highly advantageous setting for detecting the cosmological 21\,cm signal. By strictly controlling observation time within the "doubly good time" window to prevent Earth's RFI leakage and employing the VZOP algorithm, observing for about 100 days is sufficient to reliably extract the signal. However, real-world scenarios are highly complex, and the true errors in the beam are unknown. Additionally, there are standing wave reflections between the antenna and receiver, which further distort the foreground power-law spectrum. Further optimization of the antenna system for DSL is still needed in the future.

\begin{acknowledgement}
This work is supported by National Key R\&D Program of China grant no. 2022YFF0504300, the National SKA Program of China grant no. 2020SKA0110200, no. 2020SKA0110401 and the National Natural Science Foundation of China grant no. 11973047. F. Wu acknowledges the support from the National Natural Science Foundation of China grant no. 12273070. X. Chen acknowledges the support from the Chinese Academy of Science ZDKYYQ20200008 and the National Natural Science Foundation of China grant no. 12361141814. Y. Shi acknowledges the support from the National Key R\&D Program of China (2023YFA1607800, 2023YFA1607801, 2020YFC2201602), the National Science Foundation of China (11621303), CMS-CSST-2021-A02, and the Fundamental Research Funds for the Central Universities.
\end{acknowledgement}

\paragraph{Competing Interests}
None

\paragraph{Data Availability Statement}
The code and data underlying this article will be shared on reasonable request to the corresponding authors.

%\endnote in some journals will behave like \footnote; and \printendnotes will not output anything. 
\printendnotes

\printbibliography

\appendix

\section{Lunar Reflection}\label{sec:LR}
\begin{figure}
	\centering
	\includegraphics[width=\columnwidth]{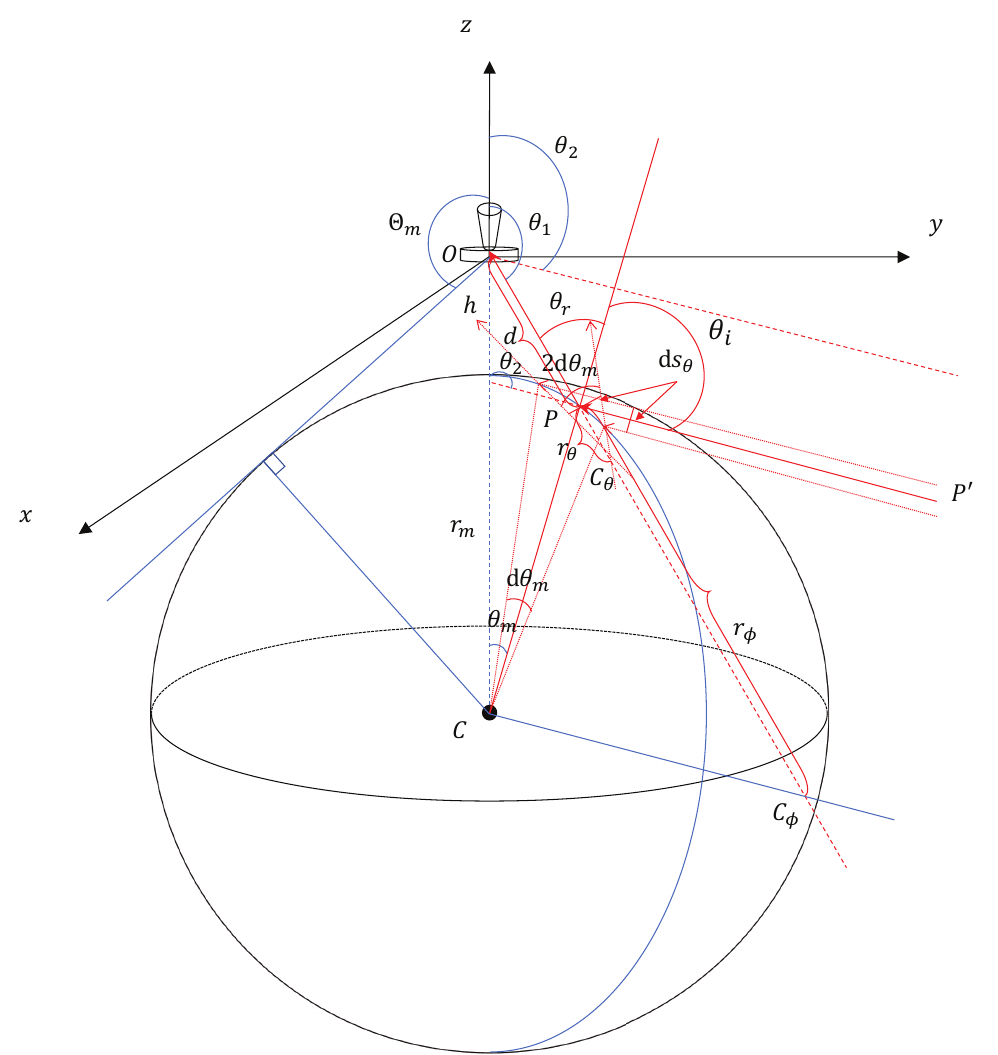}
	\caption{Schematic of Lunar reflection. The radius of the Moon is $r_m$, and the satellite is at a height $h$ above the Lunar surface. Radiation from the point source $P'$ in the infinite distance reaches the Moon, where the distance between the upper and lower radiations is $\mathrm{d}s_\theta$. These radiations undergo reflection near point $P$ on the Moon's surface. Establishing a coordinate system $(x, y, z)$ with the antenna at the centre. The angle between the line $CP$ and the $z$-axis is $\theta_m$, while the angle between the upper and lower reflection points relative to point $C$ is $\mathrm{d}\theta_m$. The backward extensions of the adjacent reflected radiations intersect at point $C_\theta$, which serves as the curvature centre. $r_\theta$ denotes the curvature radius along the $\theta$ direction. At point $P$, the corresponding chord length is also $\mathrm{d}s_\theta$, and the angle opened at $C_\theta$ is $2\mathrm{d}\theta_m$. Drawing a straight line from the infinitely distant point $P'$ to the centre of the Moon, the backward extension of the radiation after reflection at point $P$ intersect with this line at $C_\phi$. This point is the curvature centre along the $\phi$ direction, with a curvature radius of $r_\phi$. $d$ is the distance from the point $P$ to the satellite. When viewed from the satellite, the zenith angle of point $P$ is $\theta_1$, while that of point $P'$ is $\theta_2$. $\theta_i$ and $\theta_r$ represent the angles of incidence and reflection at point $P$, respectively. There is a critical angle $\Theta_m=\pi-\arcsin{[r_m/(h+r_m)]}$, with $\theta_1>\Theta_m>\theta_2$.}
	\label{fig:reflection_schematic}
\end{figure}

The schematic diagram of reflection is illustrated in Fig.~\ref{fig:reflection_schematic}. We can write the reflected sky temperature of the Lunar surface as
\begin{equation}
	T_\mathrm{refl}(\theta_1,\phi)=\mathcal{R}\mathcal{E}(\theta_1,\phi) \cdot T_b(\theta_2,\phi),
	\label{eq:reflected temperature}
\end{equation}
where the reflectance $\mathcal{R}=0.07$, and $\phi$ remains the same before and after reflection. $\mathcal{E}(\theta_1,\phi)$ is a factor that can be written as
\begin{equation}
	\mathcal{E}(\theta_1,\phi)=\mathcal{A}(\theta_1,\phi)\mathcal{B}(\theta_1,\phi),
\end{equation}
Where $\mathcal{A}(\theta_1,\phi)$ represents the effect of field of view expansion caused by spherical reflection, and $\mathcal{B}(\theta_1,\phi)$ denotes the energy attenuation effect caused by the transformation of parallel radiation into non-parallel radiation after spherical reflection. We will demonstrate later that $\mathcal{E}(\theta_1,\phi)=1$.

Firstly, we should illustrate the relationship between $\theta_1$ and $\theta_2$ from Fig.~\ref{fig:reflection_schematic}. Using the sine theorem, we obtain
\begin{equation}
	\dfrac{r_m}{\sin{(\pi-\theta_1)}}=\dfrac{h+r_m}{\sin{(\theta_1-\theta_m)}}.
	\label{eq:sine_theorem}
\end{equation}
Therefore,
\begin{equation}
	\theta_m=\theta_1+\arcsin{\left[\frac{(h+r_m)\sin(\theta_1)}{r_m}\right]}-\pi.
	\label{eq:thetam}
\end{equation}
Since parallel radiation emanates from a point $P'$ infinitely distant, the zenith angle $\theta_2$ of $P'$ observed from any location near the Moon remains constant. On the other hand, the reflection angle is equal to the incident angle. Therefore, based on geometric relationships, we can obtain
\begin{equation}
    \begin{cases}
    	\theta_i=\theta_r=\pi-\theta_1+\theta_m, \\
    	\theta_2=\theta_1-(\pi-\theta_r-\theta_i).
    	\label{eq:geometry_relationships}
    \end{cases}
\end{equation}
Solving Equation set~(\ref{eq:geometry_relationships}), we obtain
\begin{equation}
	\theta_2=\pi-\theta_1+2\theta_m.
	\label{eq:geometry_relationship}
\end{equation}
Finally, substituting equation~(\ref{eq:thetam}) into equation~(\ref{eq:geometry_relationship}), we can derive the relationship between $\theta_1$ and $\theta_2$
\begin{equation}
	\theta_2=2\arcsin{\left[\dfrac{(h+r_m)\sin{(\theta_1)}}{r_m}\right]}+\theta_1-\pi.
	\label{eq:final_relationship}
\end{equation}

The solid angle element opened by the satellite at a certain point on the Lunar surface will expand by the scaling factor $\mathcal{A}(\theta_1,\phi)$ after reflection, i.e., 
\begin{equation}
	\mathcal{A}(\theta_1,\phi)\sin{(\theta_1)}\mathrm{d}\theta_1=-\sin{(\theta_2)}\mathrm{d}\theta_2,
	\label{eq:scaling_factor}
\end{equation}
where the negative sign is because $\theta_2$ decreases as $\theta_1$ increases. According to the equation~(\ref{eq:final_relationship}),
\begin{equation}
	\mathrm{d}\theta_2=\left[1+\dfrac{2(h+r_m)\cos{(\theta_1)}}{\sqrt{r_m^2-(h+r_m)^2\sin^2{(\theta_1)}}}\right]\mathrm{d}\theta_1.
	\label{eq:differential_relationship}
\end{equation}
By combining equation~(\ref{eq:scaling_factor}) and equation~(\ref{eq:differential_relationship}), we obtain
\begin{equation}
	\mathcal{A}(\theta_1,\phi)=-\dfrac{\sin(\theta2)}{\sin{(\theta_1)}}\left[1+\dfrac{2(h+r_m)\cos{(\theta_1)}}{\sqrt{r_m^2-(h+r_m)^2\sin^2{(\theta_1)}}}\right].
	\label{eq:factor1}
\end{equation}

The radiation from a point source at infinity is collimated, and after spherical reflection, the energy diverges. Along the $\theta$ direction, three adjacent parallel beams of light undergo reflection near point $P$, with the angles between the upper and lower reflection points relative to the Lunar centre being $\mathrm{d}\theta_m$. After reflection, the three beams of light are no longer parallel, and their backward extensions intersect approximately at point $C_\theta$. Based on the characteristics of the reflection process, the angle formed by the three beams at $C_\theta$ is $2\mathrm{d}\theta_m$. Considering $C_\theta$ as the centre of curvature, the chord length $\mathrm{d}s_\theta$ at point $P$ equals the distance between the three incident light beams. The chord length between the upper and lower reflection points relative to the Lunar centre is $r_m\mathrm{d}\theta_m$. According to geometric relationships, we can derive
\begin{align}
    \mathrm{d}s_\theta
    &=r_m\mathrm{d}\theta_m\cdot\cos{(\pi+\theta_m-\theta_1)} \nonumber\\
    &=\sqrt{r_m^2-(h+r_m)^2\sin^2(\theta_1)}\mathrm{d}\theta_m.
\end{align}
The curvature radius along the $\theta$ direction is
\begin{equation}
	r_\theta=\frac{\mathrm{d}s_\theta}{2\mathrm{d}\theta_m}=\frac{\sqrt{r_m^2-(h+r_m)^2\sin^2(\theta_1)}}{2}.
	\label{eq:curvature_theta}
\end{equation}
A straight line is drawn from the source's position to the centre of the Moon, serving as an axis. All radiation reflected on a circular ring passing through point $P$ with the axis as its centre converges to a point $C_\phi$ on the axis. This is the curvature centre of the radiation reflected at point $P$ in the $\phi$ direction, with a curvature radius of
\begin{align}
    r_\phi
    &=r_m\frac{\sin(\theta_i)}{\sin(\pi-\theta_i-\theta_r)} \nonumber\\
    &=\frac{r_m^2}{2\sqrt{r_m^2-(h+r_m)^2\sin^2(\theta_1)}}.
    \label{eq:curvature_phi}
\end{align}
The distance from the reflection point to the antenna is
\begin{align}
    d
    &=r_m\frac{\sin(\theta_m)}{\sin(\pi-\theta_1)} \nonumber\\
    &=-\left[\sqrt{r_m^2-(h+r_m)^2\sin^2(\theta_1)}+(h+r_m)\cos(\theta_1)\right].
    \label{eq:distance}
\end{align}
Therefore, the energy attenuation factor is
\begin{equation}
	\mathcal{B}(\theta_1, \phi)=\frac{r_\theta r_\phi}{(d+r_\theta)(d+r_\phi)}.
	\label{eq:factor2}
\end{equation}
Substituting equation~(\ref{eq:final_relationship}) into equation~(\ref{eq:factor1}), and equations~(\ref{eq:curvature_theta}) - (\ref{eq:distance}) into equation~(\ref{eq:factor2}), we obtain
\begin{equation}
	\mathcal{E}(\theta_1,\phi)=\mathcal{A}(\theta_1,\phi)\mathcal{B}(\theta_1,\phi)=1
\end{equation}

\begin{figure}
	\includegraphics[width=\columnwidth]{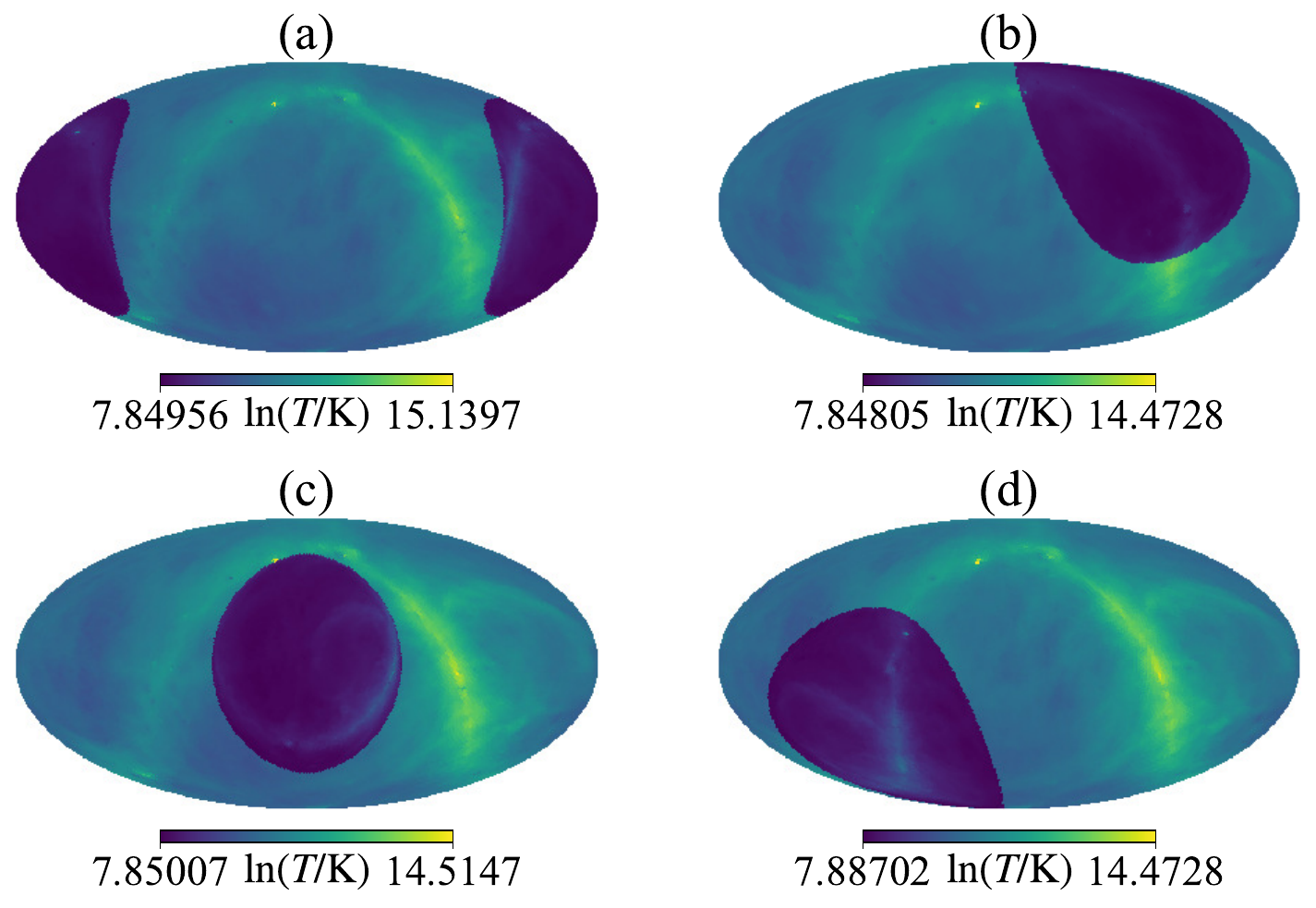}
	\caption{Sky temperature maps observed by the antenna when the satellite orbits around the Moon at $0^\circ$ (a), $90^\circ$ (b), $180^\circ$ (c), and $270^\circ$ (d).}
	\label{fig:all_moments}
\end{figure}

Fig.~\ref{fig:all_moments} presents the all-sky temperature map at different times assuming a Lunar albedo of 7\%. Panel (a) illustrates the sky temperature map when the satellite is at its initial position, while panels (b), (c), and (d) depict the sky temperature maps when the satellite has orbited the Moon by $90^\circ$, $180^\circ$, and $270^\circ$, respectively. The darker regions indicate the presence of the Moon, representing the combined effect of Lunar reflection and intrinsic blackbody temperature. All radiation from sky areas observable directly by the satellite undergoes reflection off the Moon and is subsequently observed by the antenna.

\section{Robustness Tests of VZOP}
\label{sec:robust}
\begin{figure*}
    \centering
	\includegraphics[width=\linewidth]{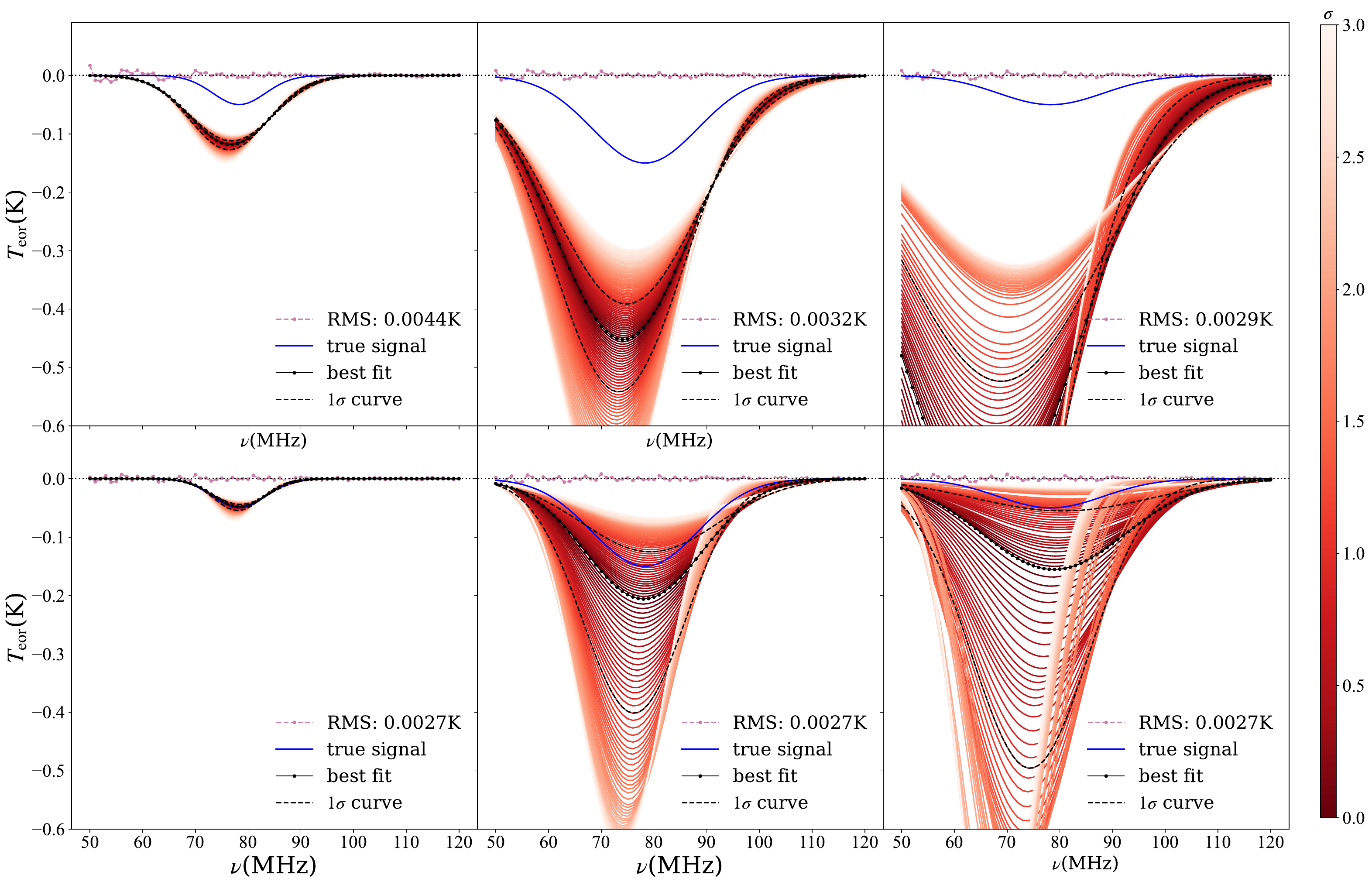}
	\caption{The fitting results of common polynomial fitting (the first row) and VZOP with 10 declination bins (the second row), when the panels from left to right correspond to different Gaussian input models. The reddish-purple dashed line shows the residuals when fitting and subtracting both the foreground and the signal. The blue solid line represents the input signal, and the black solid line is the best-fit line. Additionally, multiple lines representing different levels of errors are depicted, with varying shades of colour to indicate error magnitude, as detailed in the colour bar to the right of the figure. For clarity, two black dashed lines specifically denote the 1$\sigma$ error lines.}
	\label{fig:tests_gaussian}
\end{figure*}

All of the simulations in this section are based on the ice cream antenna. Firstly, we test the performance of VZOP and common polynomial fitting when the signal is relatively weak or broad. Using three different Gaussian models as inputs, the results of VZOP and common polynomial fitting are shown in Fig.~\ref{fig:tests_gaussian}, with the corresponding parameters listed in Table~\ref{tab:tests_gaussian}. Compared to the fiducial model discussed in Section~\ref{subsec:signal_model}, the first column represents a shallower signal, the second a broader signal, and the third a signal that is both shallower and broader. In all cases, VZOP outperforms common polynomial fitting. VZOP demonstrates robust performance even for the shallower signals. However, its effectiveness declines for broader signals, with increased uncertainties. When the signal is both shallow and broad, the performance further deteriorates. This decline is attributed to the fact that the extraction of global 21\,cm signals relies on the assumption that the foreground is significantly smoother than the signal. For broad signals, this assumption no longer holds, making signal identification more challenging. Consequently, the performance of both VZOP and common polynomial fitting suffers in such scenarios.
\begin{table}[hbt!]
\begin{threeparttable}
    \caption{Parameters of the simple models in Fig.~\ref{fig:tests_gaussian} and Fig.~\ref{fig:misidentification}. $A$: amplitude; $\nu_c$: centre frequency; $\omega$: width; $\tau$: flattening factor in the flattened Gaussian model.}
    \label{tab:tests_gaussian}
    \begin{tabular}{l|llll}
        \toprule
        \headrow & $A(\mathrm{K})$ & $\nu_c(\mathrm{MHz})$ & $\omega(\mathrm{MHz})$ & $\tau$ \\
        \midrule
        Left Column\tnote{a}  & $-0.05$ & $78.3$ & $5.0$ & - \\
        Middle Column\tnote{a} & $-0.15$ & $78.3$ & $10.0$ & - \\
        Right Column\tnote{a} & $-0.05$ & $78.3$ & $10.0$ & - \\
        \midrule
        Middle Column\tnote{b} & $-0.52$ & $78.3$ & $20.7$ & $7.0$ \\
        \bottomrule
    \end{tabular}
    \begin{tablenotes}[hang]
        \item[a] in Fig.~\ref{fig:tests_gaussian}
        \item[b] in Fig.~\ref{fig:misidentification}
    \end{tablenotes}
\end{threeparttable}
\end{table}

\begin{figure*}
    \centering
	\includegraphics[width=\linewidth]{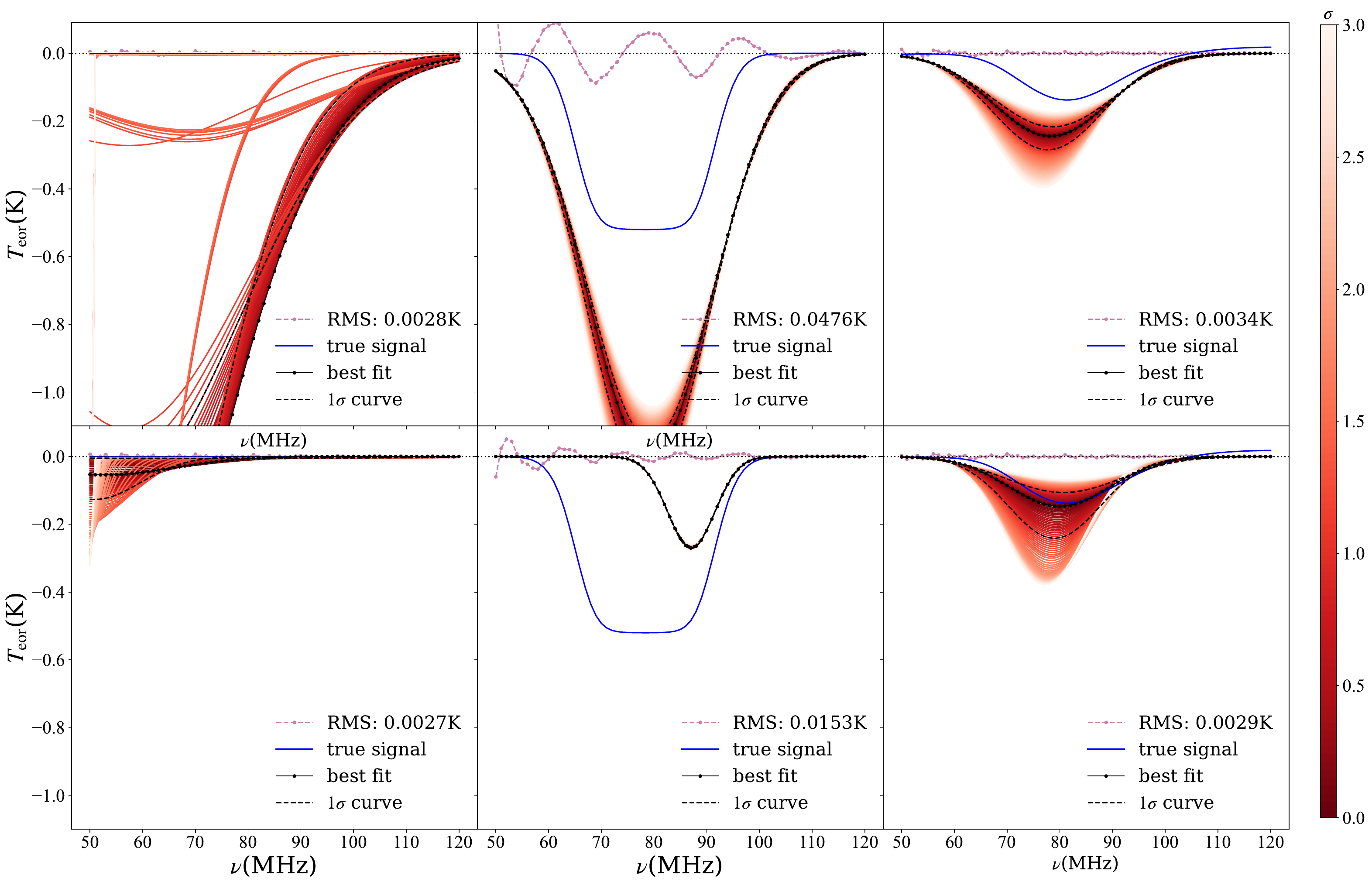}
	\caption{The fitting results of common polynomial fitting (the first row) and VZOP with 10 declination bins (the second row). The left column includes no input 21\,cm signal, the middle column uses a flattened Gaussian model detected by EDGES, and the right column uses a theoretical model generated by \texttt{globalemu}. In all cases, a Gaussian model is used for fitting. The reddish-purple dashed line shows the residuals when fitting and subtracting both the foreground and the signal. The blue solid line represents the input signal, and the black solid line is the best-fit line. Additionally, multiple lines representing different levels of errors are depicted, with varying shades of colour to indicate error magnitude, as detailed in the colour bar to the right of the figure. For clarity, two black dashed lines specifically denote the 1$\sigma$ error lines. }
	\label{fig:misidentification}
\end{figure*}

In practical observations, the fitting model may differ from the true signal. To evaluate the performance of VZOP and common polynomial fitting under such conditions, we conducted tests with results shown in Fig.~\ref{fig:misidentification}. The left column contains no input signal, yet we still apply a Gaussian model for fitting. The results demonstrate that VZOP remains largely unaffected, while common polynomial fitting becomes unstable. The middle column uses the flattened Gaussian model detected by EDGES, with parameters listed in Table~\ref{tab:tests_gaussian} \citep[][]{bowman2018absorption}. In this case, both VZOP and common polynomial fitting are misled. However, equation~(\ref{eq:thermal noise}) shows that the thermal noise should only amount to a few millikelvins, while the residuals in the figure indicate unrecognized spectral structures. The right column employs a theoretical model generated by \texttt{globalemu}, which closely resembles a Gaussian model, with parameters detailed in Table~\ref{tab:tests_theory} \citep[][]{bevins2021globalemu}. Here, VZOP outperforms common polynomial fitting. However, for most theoretical models, neither VZOP nor standard polynomial fitting succeeds in accurately identifying the signal. As discussed in Section~\ref{subsec:signal_model}, this is primarily due to the interference from the low-frequency absorption trough and high-frequency emission peak. These tests provide only a preliminary exploration of the robustness of the algorithm. In real observations, due to the complexity of the signal, a comprehensive evaluation using multiple techniques is necessary to assess the quality of models. These techniques may include Bayesian evidence, Akaike Information Criterion (AIC), Bayesian Information Criterion (BIC), Cross-Validation, and Posterior Predictive Checks, among others. These aspects will be addressed in future work.
\begin{table*}[hbt!]
\begin{threeparttable}
    \caption{Parameters of the theoretical model in the right column of Fig.~\ref{fig:misidentification}. $f_*$: star formation efficiency; $V_c$: minimal virial circular velocity; $f_X$: X-ray efficiency of sources; $\tau$: CMB optical depth; $\alpha$: power defining the slope of the X-ray SED; $\nu_\mathrm{min}$: low energy cut-off of the X-ray SED; $R_\mathrm{mfp}$: mean free path of ionizing photons.}
    \label{tab:tests_theory}
    \begin{tabular}{l|lllllll}
        \toprule
        \headrow & $f_*(\mathrm{km/s})$ & $V_c$ & $f_X$ & $\tau$ & $\alpha$ & $\nu_\mathrm{min}(\mathrm{keV})$ & $R_\mathrm{mfp}(\mathrm{Mpc})$ \\
        \midrule
        Cosmological Parameters  & $0.50$ & $52.1$ & $1.0$ & $0.070$ & $1.0$ & $0.2$ & $20.0$ \\
        \bottomrule
    \end{tabular}
\end{threeparttable}
\end{table*}

\section{Satellite Rotation}\label{sec:SR}
\begin{figure}
	\includegraphics[width=\columnwidth]{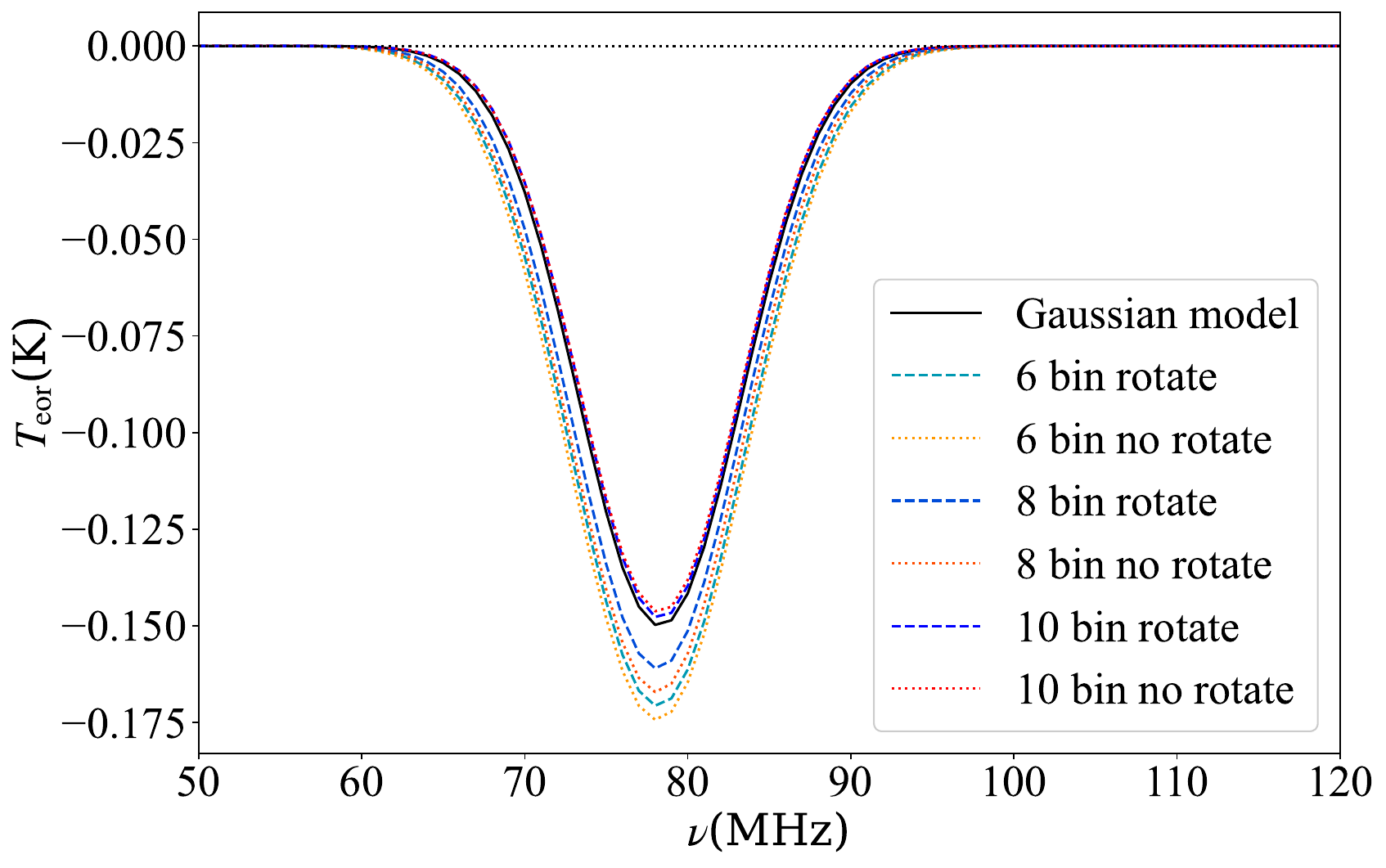}
	\caption{The satellite rotation slightly affects the fitting results when there is a 10\% error in the antenna model used by VZOP. The sky model here does not consider Lunar radiation. The cold-tone dashed lines represent the fitting results when the satellite rotates, while the warm-tone dotted lines represent the fitting results when the satellite does not rotate. The colours range from light to dark, indicating an increasing number of bins used. The solid black line provides the true signal as a reference.}
	\label{fig:rotation_necessity}
\end{figure}

Theoretically, if the antenna is non-circularly symmetric, VZOP requires satellite rotation. However, we found that when the antenna beam is precisely known, VZOP can accurately extract the 21\,cm signal even without satellite rotation, regardless of whether Lunar radiation is considered. Then we considered the non-ideal scenario, similar to Section~\ref{subsec:IABP}, assuming errors in the antenna model used by VZOP. We assume that the blade antenna has completely random errors at different pixels, but there is a correlation in the errors at different frequencies within the same pixel as described in equation~(\ref{eq:relative_error}), with an error magnitude set to 10\%. As an example, using the sky model without considering Lunar radiation, we obtained the results shown in Fig.~\ref{fig:rotation_necessity}. It can be seen that with an increase in the number of bins, the fitting accuracy gradually improves. When using 10 bins, VZOP can accurately fit the signal regardless of whether the satellite is rotating or not. However, with fewer bins, the fitting results without rotation are slightly worse than those with rotation. From the current results, it appears that rotation has minimal impact on VZOP, suggesting that satellite rotation may not be necessary. Nevertheless, due to the complexity of real-world scenarios, further evidence may be required, which will be investigated in future studies. It is important to note that the data used for the ice cream antenna model is two-dimensional, while that for the blade antenna model is three-dimensional. Therefore, even though both consider a 10\% relative error, the final error when smoothed to an average beam model differs. Hence, Fig.~\ref{fig:rotation_necessity} and Fig.~\ref{fig:antenna_errors} cannot be directly comparable.

\end{document}